\documentclass[pdftex]{sig-alternate}

\makeatletter
\def\@copyrightspace{\relax}
\makeatother

\usepackage{balance}
\usepackage[T1]{fontenc}
\usepackage[utf8]{inputenc}
\usepackage{lmodern}
\usepackage{color}
\usepackage{paralist}
\usepackage{lipsum}
\usepackage{booktabs}
\usepackage{tabularx}
\usepackage{fixltx2e}
\usepackage{microtype}
\usepackage[normalem]{ulem}
\usepackage{times}

\usepackage[]{subcaption}
\usepackage{listings}

\definecolor{mygray}{rgb}{0.5,0.5,0.5}

\usepackage[hidelinks]{hyperref}
\usepackage{xspace}

\newcommand{\ourtitle}{Learning Natural Coding Conventions}

\hypersetup{
  pdftitle={\ourtitle},
  pdfauthor={Miltiadis Allamanis, Earl T. Barr, Christian Bird, and Charles Sutton},
  plainpages=false,
  pdfpagelabels
}

\newcommand{\defref}[1]{\hyperref[#1]{Definition~\ref*{#1}}}
\newcommand{\algref}[1]{\hyperref[#1]{Algorithm~\ref*{#1}}}
\newcommand{\lineref}[1]{\hyperref[#1]{Line~\ref*{#1}}}

\newcommand{\todo}[1]{{\color{red}#1}}
\newcommand{\cb}[1]{\todo{For Chris: #1}}

\newcommand{\cut}[1]{}

\newcommand{\etal}{\hbox{\emph{et al.}}\xspace}
\newcommand{\eg}{\hbox{\emph{e.g.}}\xspace}
\newcommand{\ie}{\hbox{\emph{i.e.}}\xspace}

\newcommand{\vs}{\hbox{\emph{vs.}}\xspace}

\newcommand{\R}{\mathbb{R}}
\newcommand{\N}{\mathbb{N}}

\newcommand{\ngram}{$n$-gram\xspace}
\newcommand{\Ngram}{$N$-gram\xspace}
\newcommand{\ngrams}{$n$-grams\xspace}
\newcommand{\UNK}{\textsc{Unk}\xspace}
\newcommand{\Projname}{\textsc{Naturalize}\xspace}
\newcommand{\ProjnameT}{{\large\textbf{\textsc{Naturalize}}}\xspace}
\newcommand{\projurl}{\href{http://groups.inf.ed.ac.uk/naturalize}{\texttt{groups.inf.ed.ac.uk/naturalize}}\xspace}
\newcommand{\CP}{\mbox{\tiny \sc CP}}
\newcommand{\syn}{\mbox{\tiny \sc SYN}}

\newcommand{\junit}{\mbox{\sc JUnit}\xspace}

\newcommand{\genrule}{\texttt{genrule}\xspace} 
\newcommand{\stylish}{\texttt{stylish?}\xspace} 
\newcommand{\styleprofile}{\texttt{styleprofile}\xspace} 
\newcommand{\nateclipse}{\texttt{devstyle}\xspace} 
\newcommand{\id}[1]{\texttt{#1}} 

\usepackage{algorithmicx}
\usepackage{algpseudocode}

\makeatletter
\def\@copyrightspace{\relax}
\makeatother

\newlength{\emstr}
\newcommand{\boldpara}[1]{%
  \smallskip%
  \par\noindent\textbf{\textit{#1}}\hspace{\emstr}
}%

\newenvironment{squishlist}
{
 \begin{list}{$\bullet$}
  { \setlength{\itemsep}{0pt}
     \setlength{\parsep}{3pt}
     \setlength{\topsep}{3pt}
     \setlength{\partopsep}{0pt}
     \setlength{\leftmargin}{1.5em}
     \setlength{\labelwidth}{1em}
     \setlength{\labelsep}{0.5em} } }
{  \end{list}  }

\begin{document}

\title{\ourtitle}

\author{%
Miltiadis Allamanis$^{\dag}$
\qquad\quad Earl T. Barr${}^\ddag$
\qquad\quad Christian Bird$^{\star}$
\qquad\quad Charles Sutton$^{\dag}$\and%
\begin{tabular}{ccc}
$^{\dag}$\affaddr{School of Informatics}
& $^{\ddag}$\affaddr{Dept. of Computer Science} 
& $^{\star}$\affaddr{Microsoft Research}  \\
\affaddr{University of Edinburgh}
& \affaddr{University College London} 
& \affaddr{Microsoft} \\
\affaddr{Edinburgh, EH8 9AB, UK}
& \affaddr{London, UK} 
& \affaddr{Redmond, WA, USA} \\
\email{\{m.allamanis, csutton\}@ed.ac.uk} 
& \email{e.barr@ucl.ac.uk}
& \email{christian.bird@microsoft.com}
\end{tabular}
}
\maketitle

\begin{abstract}

Every programmer has a characteristic style, ranging from preferences about
identifier naming to preferences about object relationships and design patterns.
Coding conventions define a consistent syntactic style, 
fostering readability and hence maintainability.  When collaborating,
programmers strive to obey a project's coding conventions.
However, one third of reviews of changes contain feedback about coding conventions, 
indicating that programmers do not always follow them and that project
members care deeply about adherence.
Unfortunately, programmers are often unaware of coding conventions because inferring
them requires a global view, one that aggregates the many local
decisions programmers make and identifies emergent consensus on style.  We
present \Projname, a framework that learns the style of a codebase, and suggests
revisions to improve stylistic consistency.  \Projname builds on recent work in
applying statistical natural language processing to source code.  We apply
\Projname to suggest natural identifier names and formatting conventions. We
present four tools focused on ensuring natural code during development and
release management, including code review.  \Projname achieves
$94$\% accuracy in its top suggestions for identifier names and can even transfer
knowledge about conventions across projects, leveraging a corpus of 10,968 open
source projects.  We used \Projname to generate $18$ patches for $5$ open source 
projects: $14$ were accepted.

\end{abstract}


\section{Introduction}
\label{sec:introduction}

To program is to make a series of choices, ranging from design decisions --- like
how to decompose a problem into functions --- to the choice of identifier names and
how to format the code.  While local and syntactic, the latter are important:
names connect program source to its problem
domain~\cite{binkley:emse:13,lawrie:icpc:2006,liblit2006cognitive,takang:jpl:1996};
formatting decisions usually capture control flow~\cite{hindle:scp:2009}.
Together, naming and formatting decisions determine the readability of a
program's source code, increasing a codebase's portability, its accessibility to
newcomers, its reliability, and its maintainability~\cite[\S
1.1]{oracle:codeconventions}.  Apple's recent, infamous bug in its handling of
SSL certificates \cite{arthur2014apple,langley2014bug} exemplifies the impact that formatting can have on
reliability. Maintainability is especially important since
developers spend the majority ($80$\%) of their time maintaining code~\cite[\S
6]{swebok:2004}.  

A convention is ``an equilibrium that everyone expects in interactions that have
more than one equilibrium''~\cite{young1996economics}.  For us, coding
conventions arise out of the collision of the stylistic choices of programmers.
A \emph{coding convention} is a syntactic restriction not imposed by a programming
language's grammar.   Nonetheless, these choices are important enough that they
are enforced by software teams.
Indeed, our investigations indicate that developers enforce such coding 
conventions rigorously,
with roughly one third of code reviews containing feedback about following them
(\autoref{sec:eval:importance}).  

Like the rules of society at large, coding conventions fall
into two broad categories:  \emph{laws}, explicitly stated and enforced rules,
and \emph{mores}, unspoken common practice that emerges spontaneously.  
Mores pose a particular challenge: because 
they arise spontaneously from emergent consensus, they are inherently
difficult to codify into a fixed set of rules, so rule-based formatters
cannot enforce them, and even programmers
themselves have difficulty adhering to all of the implicit mores
of a codebase.
Furthermore, popular code changes constantly, and these changes necessarily
embody stylistic decisions, sometimes generating new conventions and sometimes
changing existing ones. To address this, we introduce the \emph{coding convention inference problem}, the
problem of automatically learning the coding conventions consistently used in a body of
source code.
Conventions are pervasive in software, ranging from preferences
about identifier names to preferences about class layout, object relationships,
and design patterns.
In this paper, we focus as a first step on local, syntactic conventions, 
namely, identifier naming and formatting. These are particularly
active topics of concern among developers, for example, almost \emph{one quarter} of 
the code reviews that we examined contained suggestions about naming.

We introduce \Projname, a framework that solves the coding convention inference
problem for local conventions,
offering suggestions to increase the stylistic consistency of a
codebase. \Projname can also be applied to infer rules for 
existing rule-based formatters. \Projname is descriptive, not
prescriptive\footnote{Prescriptivism is the attempt to specify rules for
correct style in language, \eg, Strunk and White \cite{strunk:white}.  Modern linguists
studiously avoid prescriptivist accounts, observing that many such rules are
routinely violated by noted writers.}:  
it learns what programmers actually do.
When a codebase does not reflect
consensus on a convention, \Projname recommends nothing, because it has not
learned anything with sufficient confidence to make recommendations.  
The naturalness insight of Hindle \etal~\cite{hindle2012naturalness}, building
on Gabel and Su~\cite{gabel:fse:2010}, is that most short code
utterances, like natural language utterances, are simple and repetitive.  Large
corpus statistical inference can discover and exploit this naturalness to
improve developer productivity and code robustness.  We show that coding
conventions are \emph{natural} in this sense. 


Learning from local context allows \Projname to learn syntactic restrictions, or
sub-grammars, on identifier names like camelcase or underscore, and to
\emph{unify} names used in similar contexts, which rule-based code formatters
simply cannot do.
Intuitively, \Projname works by identifying identifier names or formatting choices
that are surprising according to a probability distribution over code text.  When
surprised, \Projname determines if it is sufficiently confident to suggest a
renaming or reformatting that is less surprising;  it unifies the surprising
choice with one that is preferred in similar
contexts elsewhere in its training set.  \Projname is \emph{not} automatic; it
assists a developer, since its suggestions, both renaming and even formatting,
as in Python or Apple's aforementioned SSL bug \cite{arthur2014apple,langley2014bug}, are potentially semantically
disruptive and must be considered and approved.  \Projname's suggestions enable
a range of new tools to improve developer productivity and code quality:
    \begin{inparaenum}[1)]
      \item A pre-commit script that rejects commits that excessively 
        disrupt a codebase's conventions; 
      \item A tool that converts the inferred conventions into rules for use
        by a code formatter;
      \item An Eclipse plugin that a developer can use to check whether her 
        changes are unconventional; and
      \item A style profiler that highlights the stylistic inconsistencies
        of a code snippet for a code reviewer.
    \end{inparaenum}

\Projname draws upon a rich body of tools from statistical
natural language processing (NLP), but applies these techniques 
in a different.  NLP focuses on
\emph{understanding} and \emph{generating} language, but does not
ordinarily consider the problem of improving existing text.  The closest analog
is spelling correction, but that problem is easier because
we have strong prior knowledge about common types of spelling mistakes.
An important conceptual dimension of our suggestion problems also sets our work
apart from mainstream NLP.  In code, rare names often usefully signify unusual
functionality, and need to be preserved.  We call this the \emph{sympathetic
uniqueness principle} (SUP):  unusual names should be preserved when they appear
in unusual contexts.  We achieve this by exploiting a special token \UNK 
that is often used to represent rare words that do not appear in the
training set.  Our method incorporates SUP through a clean, straightforward
modification to the handling of \UNK.  Because of the Zipfian nature of
language, \UNK appears in unusual contexts and identifies unusual tokens that
should be preserved.  \autoref{sec:eval} demonstrates the effectiveness of this
method at preserving such names.  Additionally, handling formatting requires a
simple, but novel, method of encoding formatting.

As \Projname detects identifiers that violate code conventions and assists
in renaming, the most common refactoring~\cite{murphy2009refactor}, it is the
first tool we are aware of that uses NLP techniques to aid refactoring. 

The techniques that underlie \Projname are language independent and require only
identifying identifiers, keywords, and operators, a much easier task than
specifying grammatical structure.  Thus, \Projname is well-positioned to be
useful for domain-specific or esoteric languages for which no convention
enforcing tools exist or the increasing number of multi-language software
projects such as web applications that intermix Java, \texttt{css},
\texttt{html}, and JavaScript.

To the best of the authors' knowledge, this work is the first to address
the coding convention inference problem, to suggest names and
formatting to increase the stylistic coherence of code, and to provide tooling to
that end.  Our contributions are:

\begin{squishlist}
    
  \item We built \Projname, the first framework that presents a solution to the \emph{coding
    convention inference problem} for local conventions, including identifier
    naming and formatting, and suggests changes to increase a codebase's
    adherence to its own conventions; 

  \item We offer four tools, built on \Projname, all focused on release
    management, an under-tooled phase of the development process. 

  \item \Projname 1) achieves 94\% accuracy in its top suggestions for identifier
    names and 2) never drops below a mean accuracy of 96\% when making formatting
    suggestions; and

  \item We demonstrate that coding conventions are \emph{important} to software teams,
  by showing that 1) empirically,
  programmers enforce conventions heavily through code
  review feedback and corrective commits,  
  and 2) patches that were based on \Projname suggestions have been incorporated
  into $5$ of the most popular open source Java projects on GitHub --- of the $18$ 
  patches that we submitted, $14$  were accepted.
\end{squishlist}
\noindent Tools are available at {\footnotesize \projurl}.

\vspace*{-1ex}\section{Motivating Example} 
\label{sec:ex}

\cut{
\begin{figure}[t]
\begin{lstlisting}[xleftmargin=.9\columnsep]
public void testRunReturnsResult() {
	PrintStream oldOut = System.out;
	System.setOut(new PrintStream(
		new OutputStream() {
			@Override public void 
      write(int arg0) throws IOException {
			}
		}
	));
	try {
		TestResult result = junit.textui.
      TestRunner.run(new TestSuite());
		assertTrue(result.wasSuccessful());
	} finally {
		System.setOut(oldOut);
	}
}
\end{lstlisting}
\caption{
A snippet from \texttt{JUnit}'s \lstinline+TextRunnerTest.java+. We use this as a running example of suggesting renaming of identifiers (e.g.,  \lstinline+oldOut+, \lstinline+result+) and formatting.
}
\label{fig:ex}
\end{figure}
}

\begin{figure*}[t]
\begin{center}\resizebox{7in}{!}{
\begin{lstlisting}[xleftmargin=.7\columnsep]
public class ExecutionQueryResponse : ExecutionQueryResponseBasic<QueryResults>
{
    public ExecutionQueryResponse(QueryResults res, IReadOnlyCollection<string> str, ExecutionStepMetrics metrics) : base(res, str, metrics) { }
}
\end{lstlisting}
}
\end{center}\vspace{-12pt}
\caption{
A C\# class added by a Microsoft developer that was modified due to requests by a reviewer before it was checked in.	
}
\label{fig:msex}
\end{figure*}

\cut{
Imagine you are working on the \texttt{JUnit} project, preparing your first
contribution to an open source project.  \autoref{fig:ex} contains
\lstinline+testRunReturnResult+, a method in your patch (that also happens to be
an actual \texttt{JUnit} method in \texttt{d919bb6d}).  On line 2, you named a
\lstinline+PrintStream+ instance \lstinline{oldOut}. As good netizen, an
aspiring contributor to \texttt{JUnit}, and to avoid a tedious back and forth
about variable naming, you right-click on the name to ask \nateclipse, the
Eclipse plugin built on the \Projname framework, what it thinks of it.  Although
this name is reasonable, and may even be a good name in general, it is
inconsistent with the implicit naming convention of similar local variables
in \texttt{JUnit}, so \nateclipse suggests
\lstinline{oldPrintStream} as a better, more natural name.  Then you ask
yourself, ``What about the empty exception handler straddling lines 6--7?'':
should it contain a line break or not?  Here, again you can ask \nateclipse,
which now suggests nothing, as the line break is consonant with \texttt{JUnit}'s
style. After accepting \nateclipse's suggestions, you are confident that
your patch is consistent with the norms of the \texttt{JUnit} community, more
likely to be accepted without fuss, and send a pull request to its maintainers.
\cb{Replace this example with a real one that serves the same purpose.}
}

Both industrial and open source developers often submit their code for review
prior to check-in~\cite{rigby2013convergent}.  Consider the example of the class
shown in \autoref{fig:msex} which is part of a change submitted for review by a
Microsoft developer on February 17th, 2014.  While there is nothing
functionally wrong with the class, it violates the coding conventions of
the team.  A second developer reviewed the change and suggested that
\texttt{res} and \texttt{str} do not convey parameter meaning well enough, the
constructor line is much too long and should be wrapped.  
In the checked in change, all of these were
addressed, with the parameter names changed to \texttt{queryResults} and
\texttt{queryStrings}.

Consider a scenario in which the author had access to \Projname.  The author
might highlight the parameter names and ask \Projname to evaluate them.  At
that point it would have not only have identified \lstinline{res} and
\lstinline{str} as names that are inconsistent with the naming conventions of
parameters in the codebase, but would also have suggested better names.  The
author may have also thought to himself ``Is the constructor on line 3 too
long?'' or ``Should the empty constructor body be on it's own line and should
it have a space inside?'' Here again, \Projname would have provided immediate,
valuable answers based on the the conventions of the team.  \Projname would
indicate that the call to the base constructor should be moved to the next line
and indented
to be consonant with team conventions and that in this codebase 
empty method bodies do not need their own lines.
Furthermore it would indicate that some empty methods contain one space between the braces
while others do not, so there is no implicit convention to follow.  After
querying \Projname about his stylistic choices, the author can then be
confident that his change is consistent with the norms of the team and is more
likely to be approved during review.  Furthermore, by leveraging \Projname,
fellow project members wouldn't need to be bothered by questions about
conventions, nor would they need to provide feedback about conventions during
review.  We have observed that such scenarios occur in open source projects as
well.

\subsection{Use Cases and Tools}
\label{sec:ex:usecases}

\begin{figure}
\begin{center}
\includegraphics[width=0.9\columnwidth]{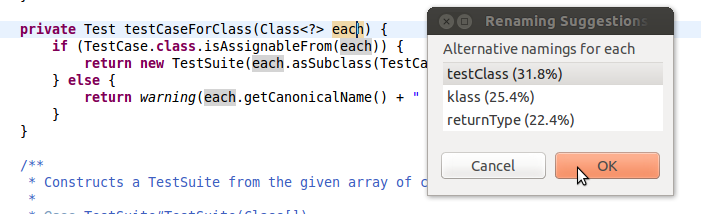}
\end{center}
\caption{A screenshot of the \nateclipse Eclipse plugin.
The user has requested suggestion for alternate names of
the \lstinline+each+ argument.\vspace{-1em}}\label{fig:screenshot}
\end{figure}

Coding conventions are critical during release management, which comprises
committing, reviewing, and promoting (including releases) changes, either
patches or branches.  This is when a coder's idiosyncratic style, isolated in
her editor during code composition, comes into contact with the styles of
others.  The outcome of this interaction strongly impacts the readability, and
therefore the maintainability, of a codebase.  Compared to other phases of the
development cycle like editing, debugging, project management, and issue
tracking, release management is under-tooled.  Code conventions are particularly 
pertinent here, and lead us to target three use cases:  1)  a
developer preparing an individual commit or branch for review or 
promotion; 2) a release engineer trying to filter out needless stylistic
diversity from the flood of changes; and 3) a reviewer wishing to consider how
well a patch or branch obeys community norms.  

Any code modification has a possibility of introducing
bugs~\cite{adams:ibm:1984,nagappan:esem:07}. This is certainly true of a
system, like \Projname, that is based on statistical inference, even when (as we
always assume) all of \Projname's suggestions are approved by a human.
Because of this risk, the gain from making a change must be worth its cost.  For
this reason, our use cases focus on times when the code is already being
changed.  To support our use cases, we have built four tools:

\begin{compactdesc}

\item[\nateclipse] A \emph{plugin for Eclipse} IDE that gives suggestions for
  identifier renaming and formatting both for a single identifier or format
  point and for the identifiers and formatting in a selection of code.

\item[\styleprofile] A \emph{code review assistant} that produces a profile that 
  summarizes the adherence of a code snippet to the coding conventions 
  of a codebase and suggests renaming and formatting 
  changes to make that snippet more stylistically consistent with a project.

\item[\genrule] A \emph{rule generator} for Eclipse's code formatter that
  generates rules for those conventions that \Projname has inferred from a
  codebase.

\item[\stylish] A high precision \emph{pre-commit script} for \texttt{Git} that rejects
  commits that have highly inconsistent and unnatural naming or formatting
  within a project.

\end{compactdesc}

The \nateclipse plugin offers two types of suggestions, single point suggestion
under the mouse pointer and multiple point suggestion via right-clicking a
selection.  A screenshot from \nateclipse is shown in \autoref{fig:screenshot}. For single point suggestions, \nateclipse displays a ranked list of
alternatives to the selected name or format.  If \nateclipse has no suggestions,
it simply flashes the current name or selection.  If the user wishes, she
selects one of the suggestions.  If it is an identifier renaming, \nateclipse
renames \emph{all} uses, within scope, of that identifier under its previous
name. This scope traversal is possible because our use cases assume an existing
and compiled codebase.  Formatting changes occur at the suggestion point.
Multiple point suggestion returns a \emph{style profile}, a ranked list of the top $k$
most stylistically surprising naming or formatting choices in the current
selection that could benefit from reconsideration.
By default, $k = 5$ based on HCI considerations~\cite{cowan2001magical,miller1956magical}.  To accept a suggestion here,
the user must first select a location to modify, then select from among its top
alternatives.  The \styleprofile tool outputs a style profile.  
\genrule (\autoref{sec:eval:genrule}) generates settings for the Eclipse code
  formatter.  Finally, \stylish is a filter that uses Eclipse code formatter
  with the settings from \genrule to 
accept or reject a commit based on its style profile.

\Projname uses an existing codebase, called a training corpus, as a reference
from which to learn conventions.  Commonly, the training corpus will be the
current codebase, so that \Projname learns domain-specific conventions related
to the projects.  Alternatively, \Projname comes with a pre-packaged suggestion
model that has been trained on a corpus of popular, vibrant projects that
presumably embody good coding conventions.  Developers can use this engine if
they wish to increase their codebase's adherence to a larger community's
consensus on best practice. Here, again, we avoid normative comparison
of coding conventions, and do not force the user to specify their
desired conventions explicitly. Instead, the user specifies a training corpus,
and this is used as an \emph{implicit} source of desired conventions.  The
\Projname framework and tools are available at \projurl.

\section{The \ProjnameT Framework}
\label{sec:framework}
  
\begin{figure*}[t]
\centering
\includegraphics[width=0.9\textwidth]{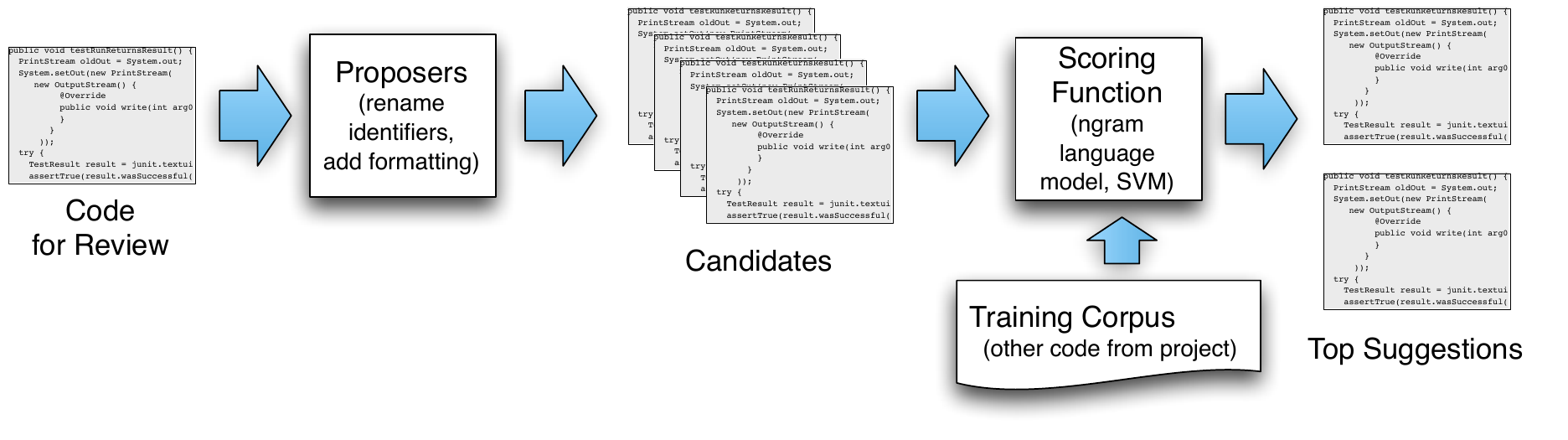}
\caption{The architecture of \Projname: a framework for learning coding 
conventions. A contiguous snippet of code is selected for review through 
the user interface. A set of \emph{proposers} returns a set of candidates,
 which are modified versions of the snippet, e.g., with one local variable renamed.
The candidates are ranked by a \emph{scoring function}, such as an \ngram language
 model, which returns a small list of top suggestions to the interface, sorted by naturalness.}
\label{fig:arch}
\end{figure*}

In this section, we introduce the generic architecture of \Projname, which can
be applied to a wide variety of different types of conventions.
\autoref{fig:arch} illustrates its architecture.  The input is a code snippet to
be naturalized.  This snippet is selected based on the user input, 
in a way that depends on the particular tool in question.  For example, in \nateclipse, if a user selects a local
variable for renaming, the
input snippet would contain all AST nodes that reference
that variable (\autoref{sec:framework:names}). The output of \Projname\ is a
short list of suggestions, which can be filtered, then presented to the
programmer.  In general, a suggestion is a set of snippets that may replace the
input snippet.  The list is ranked by a \emph{naturalness score} that is defined
below.  Alternately, the system can return a binary value indicating whether the
code is natural, so as to support applications such as \stylish.
The system makes no suggestion if it deems the input
snippet to be sufficiently natural, or is unable to find good alternatives.
This reduce the ``Clippy effect'' where users ignore a system that makes too
many bad suggestions\footnote{In extreme cases, such systems can be so widely
mocked that they are publicly disabled by the company's CEO in front of a
cheering audience: \url{http://bit.ly/pmHCwI}.}. In the next section, we describe
each element in the architecture in more detail.

\boldpara{Terminology} A language model (LM) is a probability distribution over
strings.  Given any string $x = x_0, x_1 \dots x_M$, where each $x_i$ is a
token, a LM assigns a probability $P(x).$ Let $G$ be the grammar of a
programming language.  We use $x$ to denote a snippet, that is, a string $x$
such that $\alpha x \beta \in \mathcal{L}( G )$ for some strings $\alpha,
\beta$. We primarily consider snippets that 
 that are dominated by a single node in the AST. 
We use $x$ to denote the input snippet to the framework, and $y,z$ to
denote arbitrary snippets\footnote{The application of \Projname to academic
  papers in software engineering is left to future work.}.

\subsection{The Core of \ProjnameT}
\label{sec:framework:core}

The architecture contains two main elements: proposers and the scoring function.
The \emph{proposers} modify the input code snippet to produce a list of
\emph{suggestion candidates} that can replace the
input snippet.  In the example from \autoref{fig:msex}, each candidate
replaces all occurrences of \lstinline{res} with a different name used in
similar contexts elsewhere in the project, such as \lstinline{results} or 
\lstinline{queryResults}.  In principle, many implausible suggestions
could ensue, so, in practice, proposers contain filtering logic.

A \emph{scoring function} sorts these candidates according
to a measure of naturalness. Its input is a candidate snippet,
and it returns a real number measuring naturalness.  
Naturalness is measured with respect to a training corpus 
that is provided to \Projname --- thus allowing us to follow our guiding
principle that naturalness must be measured with respect to a particular codebase.
For example,
the training corpus might be the set of source files $A$ from the current application.
A powerful way to measure the naturalness of a snippet 
is provided by statistical language modeling.
We use $P_A (y)$ to indicate the probability that
the language model $P$, which has been trained on the corpus $A$, assigns to the string $y$.
The key intuition is that an LM $P_A$ is trained so that
it assigns  high probability to strings in the training corpus, \ie,
snippets with higher log probability
are more like the training corpus, and presumably more natural.
There are several key reasons why
statistical language models are a powerful approach 
for modeling coding conventions.
First, probability distributions provide an easy way to represent
\emph{soft} constraints about conventions.
 This allows us to avoid many of the pitfalls
of inflexible, rule-based approaches.
 Second, because they are based on a learning approach,
LMs can flexibly adapt to the conventions in a new project.
Intuitively, because $P_A$ assigns high probability to strings $t \in A$ 
that occur in the training corpus, it also assigns high probability
to strings that are \emph{similar to} those in the corpus.
So the scoring function $s$ tends to favor snippets
that are stylistically consistent with the training corpus.

We score the naturalness of a snippet $y = y_{1:N}$ as
\begin{align}
\label{eq:score}
s(y, P_{A}) = \frac{1}{N} \log P_{A} (y);
\end{align}
that is, we deem snippets that are more probable under the LM
as more natural in the application $A$.
\autoref{eq:score} is cross-entropy multiplied by -$1$ to make $s$
a score, where $s(x) > s(y)$ implies $x$ is more natural than $y$.
Where it creates no confusion, 
we write $s(y)$, eliding the second argument.
When choosing between competing candidate snippets $y$ and $z$,
we need to know not only which candidate the LM prefers,
but how ``confident'' it is.
We measure this by a \emph{gap function} $g$, which is the difference in scores
%
$g(y, z, P)=s(y, P) - s(z, P).$
Because $s$ is essentially a log probability, $g$ is the log
 ratio of probabilities between $y$ and $z$.
For example, when $g(y, z) > 0$ the snippet $y$ is more natural --- \ie,
less surprising according  to the LM --- and thus
is a better suggestion candidate than $z$. If $g(y, z) =0$
then both snippets are equally natural.

\newcommand{\suggest}{\mbox{\textsf{suggest}}\xspace}
\newcommand{\suggestions}{\mbox{\textsf{suggestions}}\xspace} 
Now we define the function $\suggest(x, C, k, t)$ that returns the top 
candidates according to the scoring function. This function returns a list of top candidates,
or the empty list if no candidates are sufficiently natural.
The function takes four parameters: the input snippet $x$,
the list  $C = ( c_1, c_2, \ldots c_r)$ of candidate snippets,
and two thresholds: $k \in \N$, the maximum number of suggestions to return,
and $t \in \R,$ a minimum confidence value.  The parameter $k$ controls the size of the ranked
list that is returned to the user, while $t$ controls the \emph{suggestion frequency}, that is,
how confident \Projname needs to be before it presents any suggestions to the user.
Appropriately setting $t$ allows \Projname to avoid the Clippy effect 
by making no suggestion rather than a low quality one.
Below, we present an automated method for selecting $t$.

The \suggest function first sorts $C = (c_1, c_2, \ldots c_r)$, the candidate
list, according to $s$, so $s(c_1) \geq s(c_2) \geq \ldots \geq s(c_r)$.
Then, it trims the list to avoid overburdening the user: it truncates $C$ to
include only the top $k$ elements, so that $\mbox{length}(C) = \min\{k,r\}$.
and removes candidates $c_i \in C$ that are not sufficiently more natural than
the original snippet; formally, it removes all $c_i$ from $C$ where $g(c_i,
x) < t$.  Finally, if the original input snippet $x$ is the highest ranked in
$C$, \ie, if $c_1 = x$, \suggest ignores the other suggestions, sets $C =
\emptyset$ to decline to make a suggestion, and returns $C$.

\boldpara{Binary Decision} 
If an accept/reject decision on the input $x$ is required, \eg, as in \stylish, 
\Projname must collectively consider all of the locations in $x$ at which it could make suggestions. We propose a score function
for this binary decision that measures how good is the best possible improvement that \Projname is able to make.  Formally, let $L$ be the set of locations in $x$ at which \Projname is able to make suggestions, and for each $\ell \in L$, let $C_\ell$ be 
the system's set of suggestions at $\ell$.
In general, $C_\ell$ contains name or formatting suggestions.  Recall that $P$
is the language model.  We define the score
\begin{align}
	G(x,P) = \max_{\ell \in L} \max_{c \in C_\ell} g(c, x).
\end{align}

If $G(x,P)>T$, then \Projname rejects the snippet as being excessively
unnatural. The threshold $T$ controls the sensitivity of \Projname to unnatural
names and formatting.  As $T$ increases, fewer input snippets will be rejected,
so some unnatural snippets will slip through, but as compensation the test is
less likely to reject snippets that are in fact well-written.

\boldpara{Setting the Confidence Threshold}\label{sec:fpr}
 The thresholds in the \suggest function and the binary decision function are on log probabilities of strings, which can be difficult for
users to interpret. Fortunately these can be set automatically 
using the \emph{false positive rate (FPR)}, the proportion of snippets $x$ that in fact follow convention but that the system erroneously rejects.
We would like the FPR to be as small as possible, but unless we wish the system to make no suggestions
at all, we must accept some false positives.
So instead we set a maximum acceptable FPR $\alpha$, and we search for a
threshold $T$ that ensures
that the FPR of \Projname is at most $\alpha$.
This is a similar logic to statistical hypothesis testing.
To make this work, we need to estimate the FPR for a given $T$.
To do so, we select a random set of snippets from the training corpus,
\eg, random method bodies, and compute the proportion of the random snippets that are rejected 
using $T$. Again leveraging our assumption that our training corpus contains natural
code,
this proportion estimates the FPR. We use a grid search \cite{bergstra2012random} to find the greatest value
of $T < \alpha$, the user-specified acceptable FPR bound.

\subsection{Choices of Scoring Function}
\label{sec:framework:scoring}

The generic framework described in \autoref{sec:framework:core} 
can, in principle, employ a wide variety of machine learning or NLP methods 
for its scoring function. Indeed, a large portion of the statistical NLP literature
focuses on probability distributions over text, including language models,
probabilistic grammars, and topic models.
Very few of these models have been applied to code; exceptions include
\cite{allamanis2013mining,hindle2012naturalness,maddison2014structured,movshovitz2013natural,nguyen2013statistical}.
We choose to build on statistical language models,
because previous work of Hindle \etal.
\cite{hindle2012naturalness} has shown that they are particularly able to 
capture the naturalness of code. 

The intuition behind language modeling is that 
since there is an infinite number of possible strings,
obviously we cannot store a probability value for every one.
Different LMs make different simplifying assumptions to make the modeling tractable,
and will determine the types of coding conventions that we are able to infer.
One of the most effective practical LMs is the \ngram language model.
\Ngram models make the assumption that the next token can be
predicted using only the previous $n-1$ tokens. Formally,
the probability of a token $y_{m}$, conditioned on all of the previous tokens
$y_{1}\dots y_{m-1}$, is a function only of the previous $n-1$ tokens.
Under this assumption, we can write
\begin{equation}
\label{eq:ngramLM}
P(y_{1}\dots y_{M})=\prod_{m=1}^{M}P(y_m | y_{m-1}\dots y_{m-n+1}).
\end{equation}
To use this equation we need to know the conditional probabilities
$P(y_m | y_{m-1}\dots y_{m-n+1})$ for each possible $n$-gram.
This is a table of $V^n$ numbers, where $V$ is the number of possible
lexemes. These are the \emph{parameters} of the model
that we learn from the {training corpus}.
The simplest way to estimate the model parameters is to 
set $P(y_m | y_{m-1}\dots y_{m-n+1})$
to the proportion of times that $y_m$ follows $y_{m-1}\dots y_{m-n+1}$.
However, in practice this simple estimator does not work well, because it assumes that
$n$-grams that do not occur in the training corpus have zero probability.
Instead, $n$-gram models are trained using \emph{smoothing} methods
 \cite{chen1996empirical}.  In our work, we use Katz
smoothing.

\boldpara {Cross-Project Language Models}
As we show shortly, an \ngram model is
a good scoring function for making suggestions,
but we can do better. LMs are 
flexible and can exploit other sources of information beyond local context when 
scoring a lexeme.  For example, if
we have a global set of projects $G$ chosen for their 
good coding style,
we would like to transfer information about the
style of $G$ to the current application.
We do this by using a \emph{cross-project LM}.
We train two separate \ngram models, a global model $P_G$
trained on $G$, and a local model trained only on the current application $A$.
 Then we create a new LM $P_{\CP}$ by averaging:
\begin{align}
\label{eq:crossProjectInpt}
P_{\CP} (y) = \lambda P_G (y) + (1 - \lambda) P_A (y),
\end{align}
where $\lambda \in \lbrack 0,1 \rbrack$ is a parameter
that can be set using a development set.
The cross-project model transfers information from $G$, because for an \ngram $z$ to have high probability
according to $P_{\CP}$, it must have high probability according to both $P_G$ and $P_A$.

\boldpara{Incorporating Syntactic Context}
Our approach can also flexibly incorporate higher-level information,
such as syntactic context.
In Java, for example, type information is useful for predicting identifier names.
A simple way to exploit this is to learn a probability model over tokens based on context.
Given an AST $T$ and a location $i$ in the snippet $y$,
let $\phi(T, i)$ be a function that returns a feature vector
describing the current syntactic context, without including the identity of the token $y_i$.
For example, if $y_i$ names a variable, $\phi(T,i)$ might include the variable's type, 
determined statically via compilation.  More generally, $\phi$ could include
information about the current nesting level, whether the token is within an anonymous
inner class, and so on.
We  learn a probabilistic classifier $P_{\syn}(y_i| t_i)$ on the same
training set as for the LM.
Then we can define a new score function
\begin{align} s(y |L_v)= \log P(y) +\frac{1}{N} \sum_{i = 1}^N \log \left(P_{\syn}(y_i | \phi(T, i))\right), 
 \label{eq:framework:scoring:syntacticContext}
 \end{align}
which combines the \ngram model $P$ with the classifier $P_{\syn}$ that incorporates syntactic context.

\boldpara{Implementation}
When an \ngram model is used, we can compute the gap function $g(y,z)$
very efficiently. This is because when $g$ is used within \suggest,
ordinarily the strings $y$ and $z$ will be similar, \ie, the input snippet
and a candidate revision. The key insight is that in an \ngram model, the probability
$P(y)$ of a snippet $y = (y_1 y_2 \ldots y_N)$ 
depends only on the multiset of \ngrams that occur in $y$, that is,
\begin{align}
  \mathit{NG}(y) = \{ y_i y_{i+1} \ldots y_{i+n-1} \,|\, 0 \leq i \leq N-(n-1) \}.
\end{align}
An equivalent way to write a \ngram model is 
\begin{align}
  P(y) = \prod_{a_1 a_2 \ldots a_n \in \mathit{NG}(y)} P(a_n | a_1, a_2, \ldots a_{n-1}).
\end{align}
Since the gap function is $g(y,z) = \log \lbrack P(y) / P(z) \rbrack,$
any \ngrams	that are members both of $\mathit{NG}(y)$ and $\mathit{NG}(z)$ cancel out in the quotient,
and to compute $g$, we only need to consider those \ngrams that are not in 
$\mathit{NG}(y) \cap \mathit{NG}(z)$.
Intuitively, this means that to compute the gap function $g(y,z)$,
we need to examine the \ngrams around the locations where the snippets
$y$ and $z$ differ.
This is a very useful optimization if $y$ and $z$ are long snippets
that differ in only a few locations.

When training an LM, we take measures to deal with \emph{rare} lexemes,
since, by definition, we do not have much data about them.
We use a preprocessing step --- a common strategy in language modeling ---
that builds a vocabulary with all the identifiers that appear more
than once in the training corpus.  Let
$\text{count}(v,b)$ return the
number of appearances of token $v$ in the codebase $b$. Then, if a token has
$\text{count}(v,b) \leq 1$ we convert it to a special token, which we denote \UNK.
Then we train the \ngram model as usual.  The effect is that the \UNK
token becomes a catchall that means the model expects to see a rare token,
even though it cannot be sure which one.

\subsection{Suggesting Natural Names}
\label{sec:framework:names}

In this section, we instantiate the core \Projname framework
for the task of suggesting natural identifier names. 
We start by describing the single suggestion setting. 
For concreteness, imagine a user
of the \nateclipse plugin, who selects
an identifier and asks \nateclipse for
its top suggestions. It should be easy to see how this discussion
can be generalized to the other use cases described in \autoref{sec:ex:usecases}.
 Let $v$ be the lexeme selected by the programmer. 
 This lexeme could denote a variable, a method call, or a type. 

When a programmer binds a name to an identifier and then uses it, she implicitly
links together all the locations in which that name appears.
Let $L$ denote this set of locations, that is, 
the set of locations in the current scope in which the lexeme $v$ is used.
For example, if $v$ denotes a local variable, then $L_v$ would be the set of 
locations in which that local is used.
Now, the input snippet is constructed by finding a snippet that subsumes
all of the locations in $L_v$. Specifically, 
the input snippet is constructed by taking the lowest common ancestor in
AST of the nodes in $L_v$.

The proposers for this task retrieve a set of alternative names to $v$,
which we denote $A_v$, by retrieving other names
that have occurred in the same contexts in the training set.
To do this, for every location $\ell \in L_v$ in the snippet $x$,
we take a moving window of
length $n$ around $\ell$ and copy all the \ngrams $w_i$
that contain that token.  Call this set $C_v$ the context set, \ie,
the set of \ngrams $w_{i}$ of $x$ that contain the token $v$. 
Now we find all \ngrams in the training set that are similar to an \ngram in $C_v$
but that have some other lexeme substituted for $v$.
Formally, we set $A_v$ as the set of all lexemes $v'$ 
for which $\alpha v \beta \in C_v$ and $\alpha v' \beta$ occurs in the training set.
This guarantees that if we 
have seen a lexeme in at least one similar context, we place it in the
alternatives list. Additionally, we add to $A_{v}$ the special \UNK token;
the reason for this is explained in a moment.
Once we have constructed the set of alternative names,
the candidates are a list  $S_v$ of snippets, one for each 
$v' \in A_v$, in which all occurrences of $v$ in $x$ are
replaced with $v'$. 

The scoring function can use any model $P_{A}$, such as the \ngram model (\autoref{eq:ngramLM}).
\Ngram models work well because, intuitively, they favors names that are common
\emph{in the context} of the input snippet.
As we demonstrate in \autoref{sec:eval},
this does \emph{not} reduce to simply suggesting the most common names, such as 
 \lstinline+i+ and \lstinline+j+.  For example, suppose that the system is asked to propose
a name for \lstinline{res} in line 3 of \autoref{fig:msex}.
The \ngram model is highly unlikely to suggest \lstinline{i}, because even though the name
\lstinline{i} is common, the trigram ``\lstinline{QueryResults i ,}'' is rare.

An interesting subtlety involves names that actually \emph{should be} unique.
Identifier names have a long tail, meaning that most names are individually uncommon.
It would be undesirable to replace every rare name with common ones,
as this would violate the sympathetic uniqueness principle.
Fortunately, we can handle this issue in a subtle way:
Recall from \autoref{sec:framework:core} that, during training of the $n$-gram language model,
we convert rare names into the special \UNK token.
When we do this, \UNK exists as a token in the LM, just like any other name.
So we simply allow \Projname to return \UNK as a suggestion,
exactly the same as any other name.
Returning \UNK as a suggestion means that the model expects that
it would be natural to use a rare name in the current context.
The reason that this
preserves rare identifiers is that the \UNK token occurs in the training
corpus specifically in unusual contexts where more common names were not used.
Therefore, if the input lexeme $v$ occurs in an unusual context,
this context is more likely to match that of \UNK than of any of 
the more common tokens.

\boldpara{Multiple Point Suggestion} It is easy to adapt the
system above to the multiple point suggestion task.
Recall (\autoref{sec:ex:usecases}) that this task is to consider the set of identifiers
that occur in a region $x$ of code selected by the user, and highlight the lexemes that are least natural
in context. 
For single point suggestion, the problem is to rank different
  alternatives, \eg, different variable names, for the same code location, whereas           
 for multiple point suggestion, the problem is to rank different code locations                  against each other according to how much they would benefit from improvement.
In principle, a score function could be good at the single source problem
but bad at the multiple source problem, e.g., if the score values
have a different dynamic range when applied at different locations.

We adapt \Projname slightly to address the multiple point setting.
For all identifier names $v$ that occur in $x$, we first compute
the candidate suggestions $S_v$ as in the single suggestion case.
Then the full candidate list for the multiple point suggestion
is $S = \cup_{v \in x} S_v$; each candidate
arises from proposing a change to one name in $x$.
For the scoring function, we need to address the fact that some 
names occur more commonly in $x$ than others, and we do not want to
penalize names solely because they occur more often. So we normalize
the score according to how many times a name occurs.
Formally, a candidate $c \in S$ that has been generated by changing
a name $v$, we use the score function $s'(c) = |C_v|^{-1} s(c)$.

\subsection{Suggesting Natural Formatting}
\label{sec:framework:formatting}

\begin{figure}[t] 
\scriptsize
\begin{center}
\begin{tabular}{l|p{7cm}} 
5 &  \id{INDENT$^{3s}_{1n}$}  \lstinline+@+ \id{SPACE$^{0}$} \id{ID} \id{SPACE$^{1s}$} \lstinline+public+ \id{SPACE$^{1s}$} \lstinline+void+\\
6 &  \id{INDENT$^0_{1n}$} \id{ID} \id{SPACE$^{0}$} \lstinline+(+ \id{SPACE$^{0}$} \id{ID} \id{SPACE$^{1s}$} \id{ID} \id{SPACE$^{0}$} \lstinline+)+ \id{SPACE$^{1s}$}
		\lstinline+throws+ \id{SPACE$^{1s}$} \id{ID} \id{SPACE$^{1s}$} \lstinline+{+\\
7 &  \id{\uline{INDENT}$^0_{1n}$} \id{\}} \\
8 &  \id{INDENT$^{-3s}_{1n}$} \id{\}}
\end{tabular}
\end{center}
\normalsize{}

\caption{The formatting tokenization of lines from a snippet of code 
of \texttt{TextRunnerTest.java} in \texttt{JUnit}.}
\label{fig:formattingTokenization}
\end{figure}

We apply \Projname to build a \emph{language-agnostic code formatting suggester}
that automatically and adaptively learns formatting conventions and generates
rule for use by code formatters, like the Eclipse formatter.  \Ngram models work
over token streams; for the \ngram instantiation of \Projname to provide
formatting suggestions, we must convert whitespace into tokens. We change the
tokenizer to encode \emph{contiguous} whitespace into tokens
using the grammar

\resizebox{0.7\columnwidth}{!}{
\noindent\begin{tabular}{ll}
  & \id{S} ::= \id{T W S} | $\epsilon$ \\
  & \id{W} ::= \id{SPACE$^{space/tab}$} | \id{INDENT$^{space/tabs}_{lines}$} \\
  & \id{T} ::= \id{ID} | \id{LIT} | \id{\{} | \id{\}} | \id{.} | \id{(} | \id{)}
  | \id{<keywords>}.
\end{tabular}
}

\noindent\autoref{fig:formattingTokenization} shows a sample tokenization of a
code snippet drawn from \texttt{JUnit}.  We collapse all identifiers to a
single \texttt{ID} token and all literals to a single \texttt{LIT} token because
we presume that the actual identifier and literal lexemes do not convey
information about formatting.  Whitespace at the start of a line determines the
indentation level, which usually signifies nesting.  We replace it with
a special tokens \texttt{INDENT}, along with metadata encoding the 
increase of whitespace (that may be negative or zero) \emph{relative} to
the previous line. We also annotate \id{INDENT} with the number of new 
lines before any proceeding (non-whitespace) token. This captures
code that is visually separated with at least one empty line.
In \autoref{fig:formattingTokenization},
line 5 indents by 3 spaces in the directly next line.  Whitespace
within a line, controls the appearance of operators and punctuation:
``\lstinline+if(+ \vs \lstinline+if (+'' and ``\lstinline+x<y+ \vs 
\lstinline+x < y+''.  We encode this whitespace into the special token 
\texttt{SPACE}, along the number of spaces/tabs that it contains. If 
between two non-whitespace tokens there is no space we add a
\texttt{SPACE$^0$}. Finally, we annotate all tokens of type \id{T} with
their size (number of characters) and the current column
of that token. This helps us capture the increasing probability of
an \id{INTENT} when we reach a large line length. To reduce sparsity
we quantize these two annotations into buckets of size $q$.

Although hard to see, an empty exception handler is straddling lines 6--7 in
\autoref{fig:formattingTokenization}.  If a programmer asks if this is
convention, \Projname considers alternatives for the underlined token.  We train
the LM over the whitespace-modified tokenizer, then, since the vocabulary of
whitespace tokens (assuming bounded numbers in the metadata) is small, we rank
all whitespace token alternatives according to the scoring function
(\autoref{sec:framework:scoring}).  

\subsection{Converting Conventions into Rules} 
\label{sec:eval:genrule}

\Projname can convert the conventions it infers from a codebase into rules for a
code formatter.  We formalize a code formatter's rule as the set of settings $S
= \{s_1,s_2,\dots,s_n\}$ and $C$, a set of constraints over the elements in $S$.
For example $s_i$ could be a boolean that denotes ``\lstinline+{+ must be on the
  same line as its function signature'' and $s_j$ might be the number of spaces
  between the closing \lstinline+)+ and the \lstinline+{+.  Then $C$ might
    contain $(s_j \geq 0) \rightarrow s_i \wedge (s_j < 0) \rightarrow \neg
    s_i$.  To extract rules, we handcraft a set of minimal code snippets that
    exhibit each different setting of $s_i$, $\forall s_i \in S$.  After training \Projname
    on a codebase, we apply it to these snippets.  Whenever \Projname is
    confident enough to prefer one to the other, we infer the appropriate
    setting, otherwise we leave the default untouched, which may be to ignore
    that setting when applying the formatter.

\begin{figure*}[t]
\begin{center}
	\footnotesize
\begin{tabular}{lrrrlp{5cm}} \hline
Name & Forks & Watchers & Commit &Pull Request& Description \\ \hline
\textsf{elasticsearch} & 1009 & 4448 & \texttt{af17ae55} &$\href{https://github.com/elasticsearch/elasticsearch/pull/5075}{\texttt{\#5075}^\text{merged}}$& REST Search Engine\\
\textsf{libgdx} &  1218 & 1470 & \texttt{a42779e9}&$\href{https://github.com/libgdx/libgdx/pull/1400}{\texttt{\#1400}^\text{merged}}$& Game Development Framework\\
\textsf{netty} &  683 & 1940 & \texttt{48eb73f9} &$^\text{did not submit}$&Network Application Framework\\
\textsf{platform\_frameworks\_base} &  947 & 1051 &\texttt{a0b320a6}&$^\text{did not submit}$& Android Base Framework\\
\textsf{junit}* &  509 & 1796 & \texttt{d919bb6d}&$\href{https://github.com/junit-team/junit/pull/834}{\texttt{\#834}^\text{merged}}$&Testing Framework\\
\textsf{wildfly} &  845 & 885 & \texttt{9d184cd0}&$^\text{did not submit}$&JBoss Application Server\\
\textsf{hudson} &  997 & 215 &\texttt{be1f8f91}&$^\text{did not submit}$&Continuous Integration Server\\
\textsf{android-bootstrap} &  360 & 1446& \texttt{e2cde337}&$^\text{did not submit}$&Android Application Template\\
\textsf{k-9} &  583 & 960 &\texttt{d8030eaa}&$\href{https://github.com/k9mail/k-9/pull/454}{\texttt{\#454}^\text{merged}}$&Android Email Client \\
\textsf{android-menudrawer} &  422 & 1138 &\texttt{96cdcdcc}&$\href{https://github.com/SimonVT/android-menudrawer/pull/216}{\texttt{\#216}^\text{open}}$&Android Menu Implementation\\ \hline
\end{tabular}
\\{\scriptsize *Used as a validation project for tuning parameters.}
\end{center}
\normalsize{}
\caption{Open-source Java projects used for evaluation. Ordered by popularity.}
\label{tbl:projects}
\end{figure*}

\section{Evaluation}
\label{sec:eval}

We now present an evaluation of the value and effectiveness of \Projname.  This
evaluation first presents two empirical studies that show \Projname solves a
real world problem that programmers care about (\autoref{sec:eval:importance}).
These studies demonstrate that
1) programmers do not always adhere to coding conventions and yet 2) that
project members care enough about them to correct such violations.
Then, we move on to evaluating the suggestions produced by \Projname.
We perform an extensive automatic evaluation (\autoref{sec:eval:suggestion}) 
which verifies that \Projname produces natural suggestions that matches real code.
Automatic evaluation is a standard methodology in statistical NLP \cite{bleu,rouge},
and is a vital step when introducing a new research problem, because it allows
future researchers to test new ideas rapidly.
This evaluation relies on perturbation:  given code text, we perturb its
identifiers or formatting, then check if \Projname suggests the original name or
formatting that was used. Furthermore, we also employ the automatic evaluation
to show that \Projname is robust to low quality corpora (\autoref{sec:eval:robustness}).

Finally, to complement the automatic evaluation, we perform
two qualitative evaluations of the effectiveness of \Projname suggestions.
First, we manually examine the output of \Projname, showing
that even high quality projects contain many entities
for which other names can be reasonably considered (\autoref{sec:eval:needless}).
Finally, we submitted patches based on \Projname suggestions (\autoref{sec:eval:patches}) to $5$
of the most popular open source projects on GitHub --- of the $18$ patches 
that we submitted, $12$ were accepted.

\boldpara{Methodology} Our corpus is a set of well-known open source Java
projects. From GitHub\footnote{\url{http://www.github.com}, on 21 August,
2013.}, we obtained the list of all Java projects that are not forks and scored
them based on their number of ``watchers'' and forks. The mean number of
watchers and forks differ, so, to combine them, we assumed these numbers follow
the normal distribution and summed their z-scores. For these evaluations 
reported here, we picked the top 10 scoring projects that are
\emph{not} in the training set of the GitHub Java Corpus \cite{allamanis2013mining}. Our original
intention was to also demonstrate cross-project learning, but have no
space to report these finding.
\autoref{tbl:projects} shows the selected projects.

Like any experimental evaluation, our results are not immune to the standard
threat to external validity that, if poorly chosen, the evaluation corpus may
not be representative of Java programs, let alone programs in other languages.
Interestingly, this threat does not extend to the training corpus, because the
whole point is to bias \Projname toward the conventions that govern the training
set.  Rather, our interest is to ensure that \Projname's performance on our
evaluation corpus matches that of projects overall, which is why we took such
care in constructing our corpus.

Our evaluations use leave-one-out cross validation. We test on each file in the
project, training our models on the remaining files. This reflects the usage
scenario that we recommend in practice. We report the average performance
over all test files. 
 For an LM, we have used a $5$-gram model, chosen via
calibration on the validation project
\texttt{JUnit}. We picked \texttt{JUnit} as the validation project because of
its medium size.

\subsection{The Importance of Coding Conventions}
\label{sec:eval:importance}

To assess whether obeying coding conventions, specifically following
formatting and naming conventions, is important to software teams today, we
conducted two empirical studies that we present in this section.  
But first, we posit that coding style is both an important and a contentious
topic. The fact that many languages and projects have style guides is a
testament to this assertion.  For example, we found that the Ruby style guide
has at least $167$ un-merged forks and the Java GitHub Corpus
\cite{allamanis2013mining} has $349$ different \texttt{.xml} configurations for
the Eclipse formatter. 

\boldpara{Commit Messages}
We manually examined $1,000$ commit messages drawn randomly from the commits of
eight popular open source projects looking for mentions of renaming, changing
formatting, and following other code conventions.  We found that 2\% of changes
contained formatting improvements, 1\% contained renamings, and 4\% contained
any changes to follow code conventions (which include formatting and renaming).
We observed that not all commits that contain changes to adhere to
conventions mention such conventions in the commit messages. Thus, our
percentages likely represent lower bounds on the frequency of commits that
change code to adhere to conventions.

\boldpara{Code Review Discussions}  
We also examined discussions that
occurred in reviews of source code changes.  Code review
is practiced heavily at Microsoft in an effort to ensure that changes are free of defects and
adhere to team standards.  Once an author has completed a change, he creates
a code review and sends it to other developers for
review.  They then inspect the change, offer feedback, and either sign off or
wait for the author to address their feedback in a subsequent change.  As part
of this process, the reviewers can highlight portions of the code and
begin a discussion (thread) regarding parts of the change (for more details
regarding the process and tools used, see Bacchelli
\etal~\cite{bacchelli2013eoc}). 

We examined $169$ code reviews selected randomly across Microsoft product groups
during 2014.  Our goal was to include enough reviews to examine at
least $1,000$ discussion threads.  In total, these $169$ reviews contained 1093
threads.  We examined each thread to determine if it contained feedback related
to a) code conventions in general, b) identifier naming, and c) code
formatting.    
$18$\% of the threads examined provided feedback regarding coding conventions of
some kind.  $9$\% suggested improvements in naming and $2$\% suggested changes
related to code formatting (subsets of the $18\%$).  In terms of the reviews that
contained feedback of each kind, the proportions are $38$\%, $24$\%,
and $9$\%.  

During February 2014, just over $126,000$ reviews were completed at Microsoft.
Thus, based on confidence intervals of these proportions,
between $7,560$ and $18,900$ reviews received feedback regarding
formatting changes that were needed prior to check-in and between $21,420$ and
$39,060$ reviews resulted in name changes in just one month.  

\begin{figure}
\resizebox{1.0\columnwidth}{!}{
\centering
\begin{tabular}{lrcrcc}
\toprule
Type & \multicolumn{2}{c}{Reviews (CI)} & \multicolumn{2}{c}{Commits (CI)} & $p-val$\\
\midrule
Conventions & 38\% & (31\%--46\%) & 4\% & (3\%--6\%) & $p \ll 0.01$ \\
Naming & 24\% & (17\%--31\%) & 1\% & (1\%--2\%) & $p \ll 0.01$ \\
Formatting & 9\% & \ \ (6\%--15\%) & 2\% & (1\%--3\%) & $p \ll 0.01$ \\
\bottomrule
\end{tabular}
}
\caption{Percent commits with log messages and reviews that contained feedback regarding
code conventions, identifier naming, and formatting with 95\% confidence intervals
in parentheses.}
\label{tbl:studies}
\end{figure}

\autoref{tbl:studies} summarizes our findings from examining commit messages and
code reviews.  We also present $95$\% confidence intervals based on the sampled
results~\cite{dowdy2011statistics}.  These results demonstrate that changes, to
a nontrivial degree, violate coding conventions even after the authors consider
them complete and also that team members expect that these violations be fixed.
We posit that, like defects, many convention violations die off during the
lifecycle of a change, so that few survive to review and fewer still escape into
the repository.  This is because, like defects, programmers notice and fix many
violations themselves during development, prior to review, so reviewers must
hunt for violations in a smaller set, and committed changes contain still fewer,
although this number is nontrivial, as we show in \autoref{sec:eval:needless}.
Corrections during development are unobservable.  However, we can compare
convention corrections in review to corrections after commit.  We used a
one-sided proportional test to evaluate if more coding conventions are corrected
during review than after commit.  The last column in \autoref{tbl:studies}
contains the p-values for our tests, indicating that the null hypothesis can be
rejected with statistically significant support.

\subsection{Suggestion}
\label{sec:eval:suggestion}

\begin{figure*}[tb]
	\centering
	\begin{subfigure}[b]{0.30\textwidth}
		\centering
		\includegraphics[width=\textwidth]{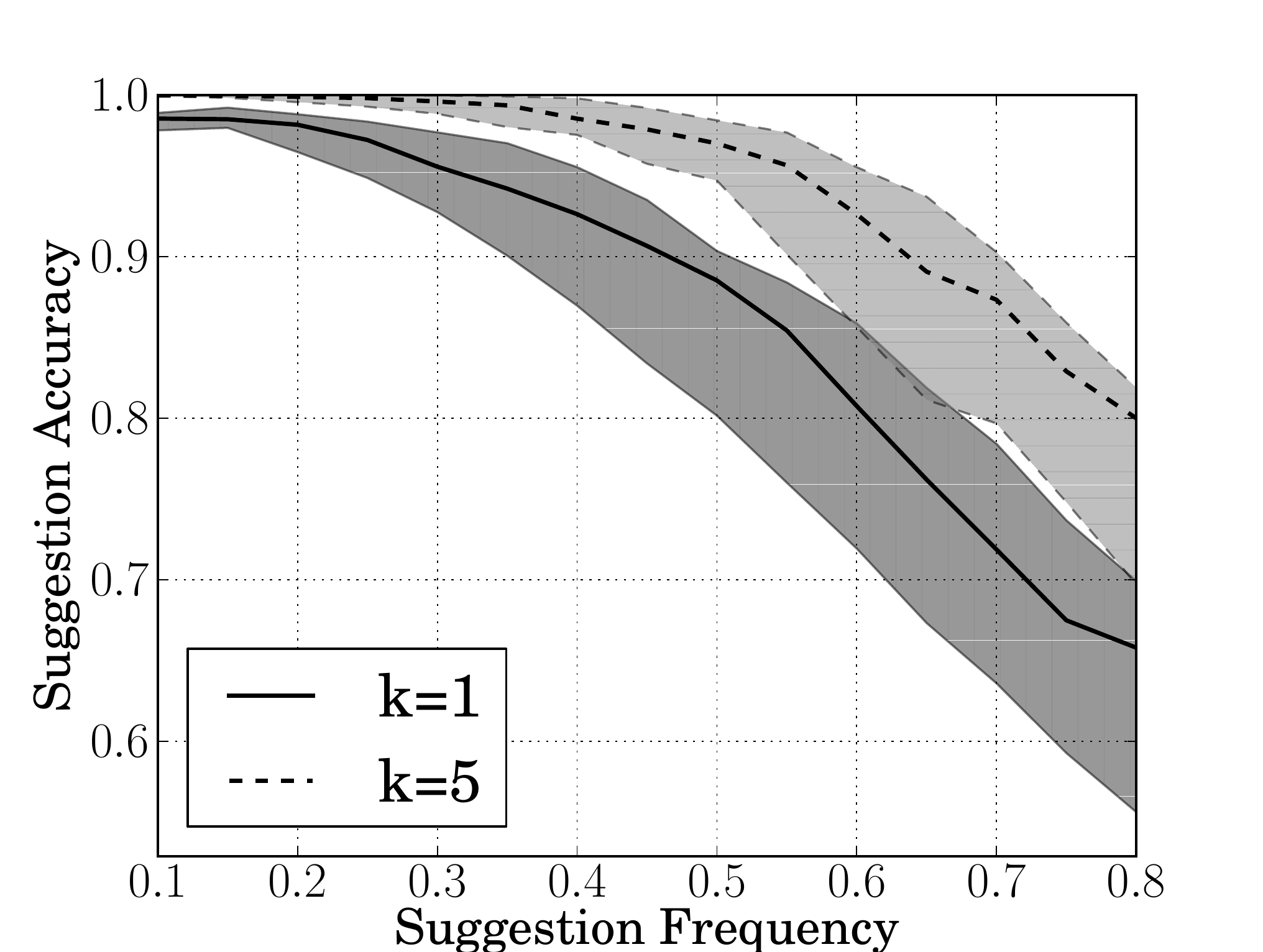}
		\caption{Variables}
		\label{fig:point:vars}		
	\end{subfigure}
\qquad
	\begin{subfigure}[b]{0.30\textwidth}
		\centering
		\includegraphics[width=\textwidth]{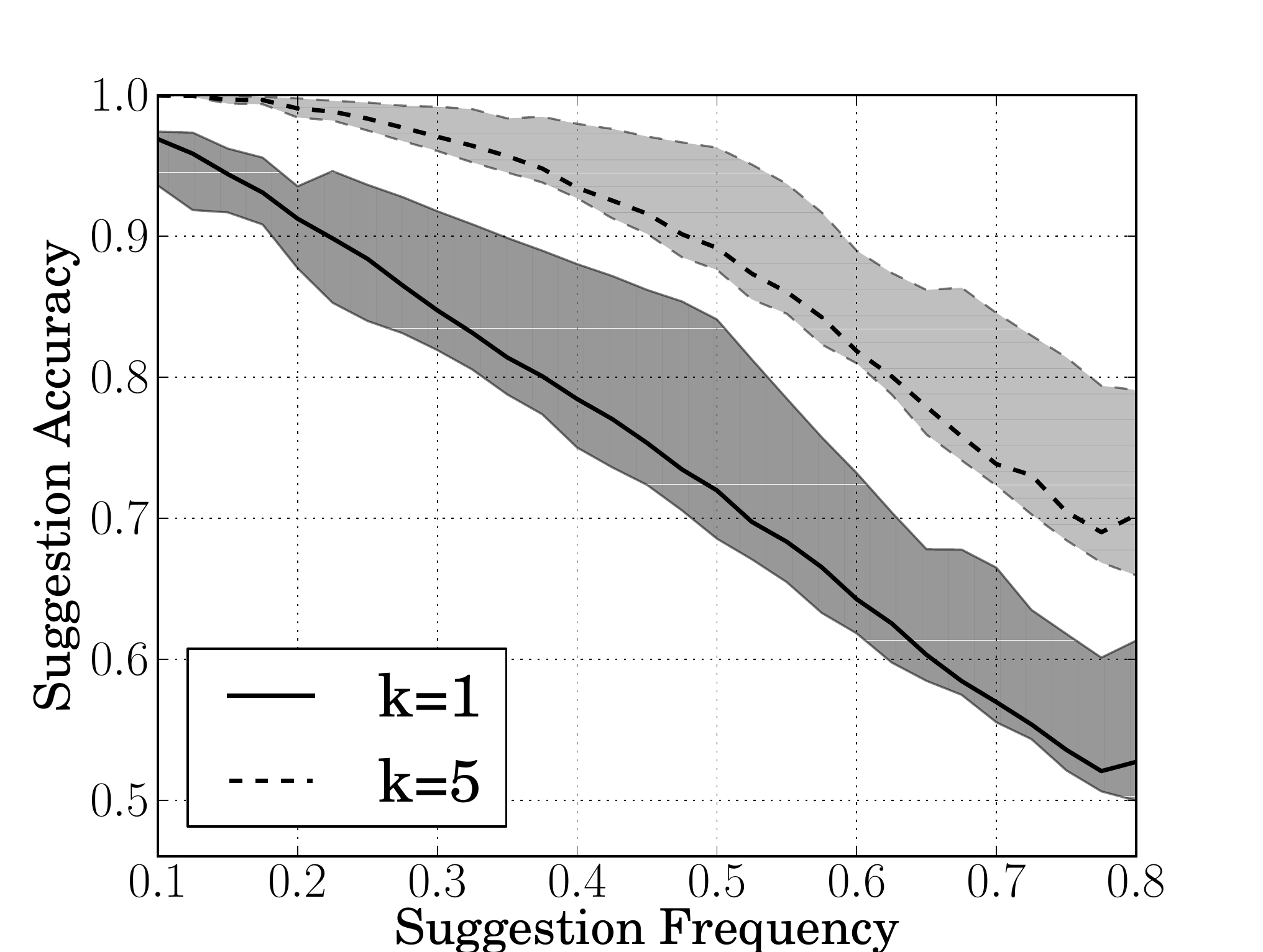}
		\caption{Method Calls}
		\label{fig:pointMethodEval}		
	\end{subfigure}
\qquad
	\begin{subfigure}[b]{0.30\textwidth}
		\centering
		\includegraphics[width=\textwidth]{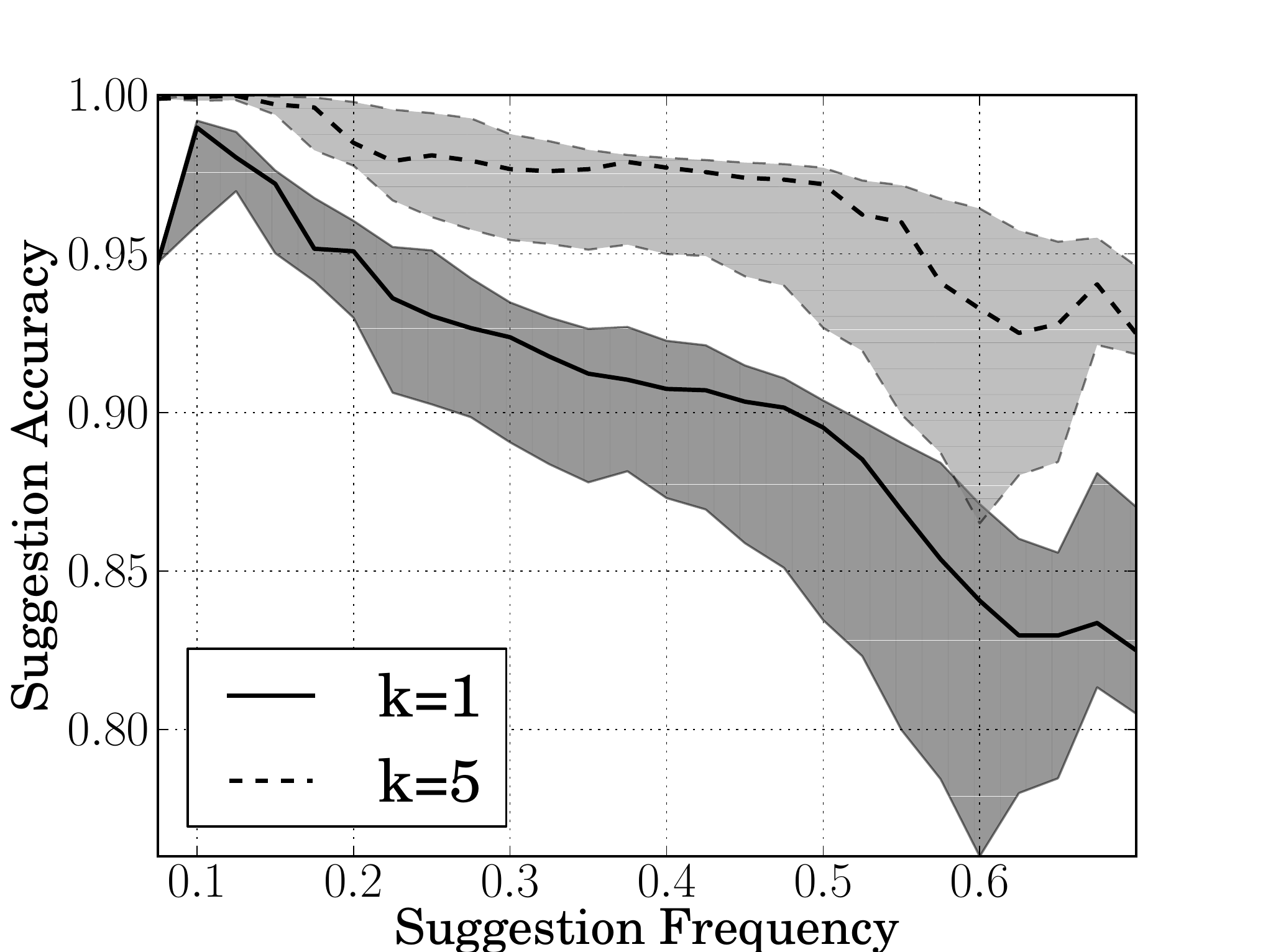}
		\caption{Typenames}
		\label{fig:pointTypeEval}		
	\end{subfigure}
	\caption{Evaluation of single point suggestions of \Projname, when it is allowed
	to suggest $k=1$ and $k=5$ alternatives.
	The shaded area around each curve shows the interquartile range
	of the suggestion accuracy across the 10 evaluation projects.}
	\label{fig:pointEval}
\end{figure*}

In this section we present an automatic evaluation of \Projname's suggestion accuracy.
First we evaluate naming suggestions (\autoref{sec:framework:names}).
We focus on suggesting new names for (1) locals, (2) arguments, (3) fields, (4) method calls, and (5)
types (class names, primitive types, and enums) --- these are the five distinct
types of identifiers the Eclipse compiler recognizes~\cite{eclipseJdt}.
Recall from \autoref{sec:framework:names} that when \Projname suggests a renaming,
it renames \emph{all} locations where that identifier is used at once.
Furthermore, as described earlier, 
we always use leave-one-out cross validation, so 
we never train the language model on the files for which we are making suggestions. 
Therefore, \Projname cannot pick up the correct name for an identifier 
from other occurrences in the same file; instead, it must generalize by learning
conventions from other files in the project.


\boldpara{Single Point Suggestion}
First we evaluate \Projname on the single point suggestion task, that is,
when the user has asked for naming suggestions for a single identifier.
To do this, for each test file, for each unique identifier
we collect {all} of the locations where the identifier occurs
 and names the same entity,
and ask \Projname to suggest a new name, renaming all occurrences at once.
We measure accuracy, that is, the percentage of the time that \Projname
correctly suggests the original name.
This is designed to reflect the typical usage scenario described in \autoref{sec:ex},
in which a developer has made a good faith effort to follow a project's
conventions, but may have made a few mistakes.

\autoref{fig:pointEval} reports on the quality of the suggestions.
Each point on these curves corresponds to a different value of the confidence
threshold $t$.
The $x$-axis shows the suggestion frequency, i.e., at what proportion of code locations
where \Projname is capable of making a suggestion does it choose to do so.
The $y$-axis shows suggestion accuracy, that is,
the frequency at which the true name is found in the top
$k$ suggestions, for $k=1$ and $k=5$.
As $t$ increases, \Projname makes fewer suggestions of higher quality,
so frequency decreases as accuracy increases.
These plots are similar in spirit to precision-recall curves in that curves
nearer the top right corner of the graph are better.
\autoref{fig:point:vars}, \autoref{fig:pointMethodEval}, and
\autoref{fig:pointTypeEval} show that \Projname performance varies with both
project and the type of identifiers.  \autoref{fig:point:vars} combines locals,
fields, and arguments because their performance is similar.  \Projname's
performance varies across these three categories of identifiers because of
the data hungriness of \ngrams and because local context is an imperfect proxy
for type constraints or function semantics.  The results show that \Projname effectively
avoids the Clippy effect, because by allowing the system to decline to suggest in a relatively small
proportion of cases, it is possible to obtain good suggestion accuracy. 
 Indeed, \Projname can achieve 94\% suggestion accuracy across identifier types,
even when forced to make suggestions at half of the possible opportunities.

\begin{figure}[tb]
	\centering
	\includegraphics[width=0.95\columnwidth]{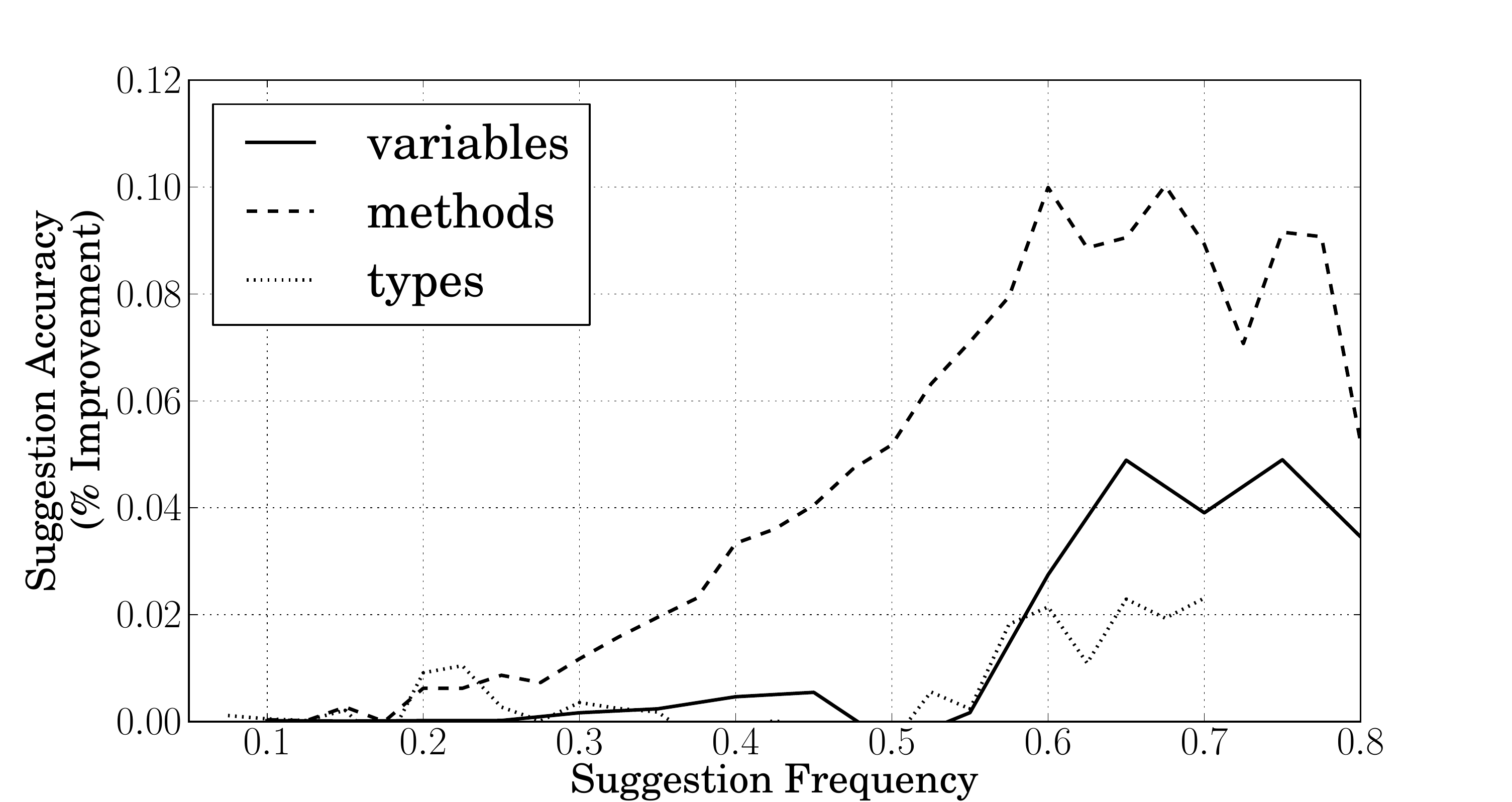}
		\caption{Improvement of suggestion accuracy using cross-project 
		  \ngram model at $k=5$}
	\label{fig:point:interpolated}
\end{figure}

\boldpara{Augmented Models}
Cross-project LMs raise the opportunity to exploit
information across projects to improve \Projname's performance.
Indeed, across our corpus,
we find that on average 14.8\% of variable names are shared
across projects. This is an indication that we can transfer useful information
about identifier naming across projects. The cross-project model
(\autoref{sec:framework:scoring}) is an approach to achieving that.
\autoref{fig:point:interpolated} shows the percent improvement in suggestion accuracy at $k=5$
when using the cross-project scoring model versus the base model.
Our evaluation suggest that the cross-project scoring model 
yields improvements of about 3\% for variables and types and an
improvements of about 10\% for method calls at the suggestion frequency of 
60\%, indicating that transferring information is feasible. We note that
for smaller suggestion frequencies this improvement diminishes.
We also experimented with incorporating syntactic context into \Projname, as
shown in \autoref{eq:framework:scoring:syntacticContext}.  Our initial results
find this strategy yields minor performance improvements in single variable
renaming suggestions with the best improvement being 2\% in the suggestion
accuracy of field names.  We believe the problem is one of finding proper
weighting for the priors, which we leave to future work.

\boldpara{Multiple Point Selection}
To evaluate \Projname's accuracy at multiple point suggestion, e.g., in \nateclipse
or \styleprofile, we mimic code snippets in which one name violates
the project's conventions. For each test snippet, we randomly choose one identifier 
and perturb it to a name 
that does not occur in the project, compute the style profile, and measure
 where the perturbed name appears in the list of suggestions.  \Projname's recall at rank $k = 7$, chosen because humans can take in
$7$ items at a glance~\cite{cowan2001magical, mays91context} is 64.2\%.  The
mean reciprocal rank is $0.47$:  meaning that, on average, we
return the bad name at position $2$. 

\begin{figure}[tb]
	\centering
	\includegraphics[width=.9\columnwidth]{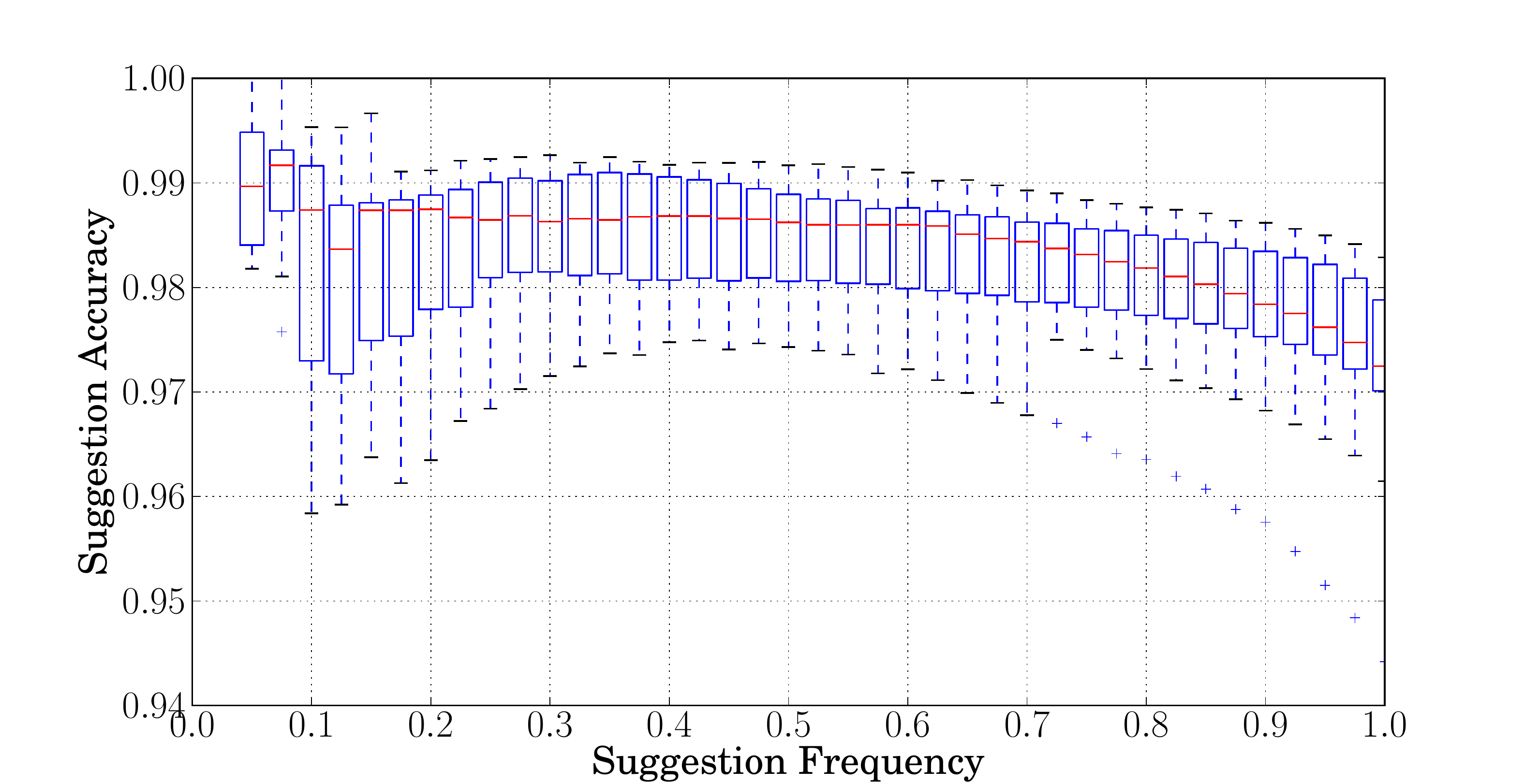}
	\caption{Evaluation of single point evaluation for formatting. 
	Only $k=1$ suggestions are allowed in the ranked list.
	The boxplots show the variance in performance across the
	10 evaluation projects.}
	\label{fig:formattingPoint}
\end{figure}

\boldpara{Single Point Suggestion for Formatting}
To evaluate \Projname's performance at making formatting suggestions,
we follow the same procedure as
the single-point naming experiment to check if \Projname correctly recovers the
original formatting from the context.
We train a
$5$-gram language model using the modified token stream ($q=20$) discussed in
\autoref{sec:framework:formatting}. We allow the system to make only $k=1$ suggestions
to simplify the UI.  We find that the system is extremely effective (\autoref{fig:formattingPoint}) at formatting suggestions,
achieving 98\% suggestion accuracy even when it is required to reformat half
of the whitespace in the test file.  This is remarkable for a system
that is not provided with any hand-designed rules about what formatting is desired.
Obviously if the goal is to reformat \emph{all} the whitespace an entire file, a rule-based formatter is  
called for.  But this performance is more than high enough to support our use cases,
such as providing single point formatting suggestions on demand, rejecting
snippets with unnatural formatting and extracting high confidence rules
for rule-based formatters.

\boldpara{Binary Snippet Decisions}
Finally, we evaluate the ability of \stylish to
discriminate between code selections that follow conventions well
from those that do not, by mimicking commits that contain
unconventional names or formatting. Uniformly at random, we selected a
set of methods from each project ($500$ in total), then with probability uniform
probability ($\frac{1}{3}$) we either made no changes or perturbed one 
identifier or whitespace token to a token in the \ngram's vocabulary $V$. 
We argue that 
this method for mimicking commits is probably a worst case for our method, because the perturbed methods
will be very similar to existing methods, which we presume to follow conventions
most of the time. 

\begin{figure}[tb]
	\centering
	\includegraphics[width=.75\columnwidth,height=.45\columnwidth,keepaspectratio=false]{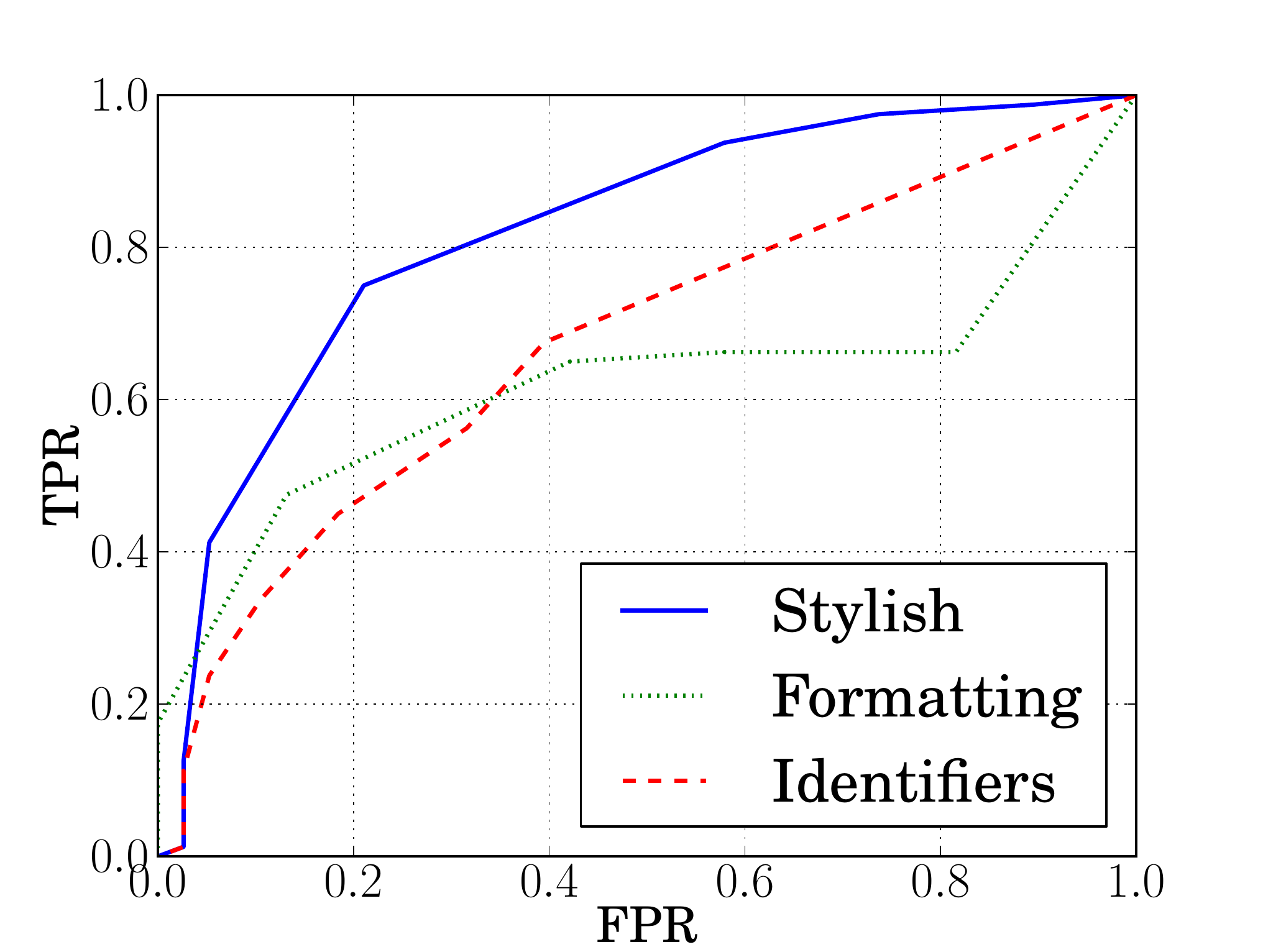}
	\caption{Evaluation of \stylish tool for rejecting unnatural changes.
      	 To generate unnatural code, we perturb one identifier or formatting point or make no changes, and evaluate whether \Projname correctly
	rejects or accepts the snippet. The graph shows the receiver 
	operating characteristic (ROC) of this process for \stylish when 
	using only identifiers, only formatting or both.}
	\label{fig:robustness}
\end{figure}

We run \stylish
and record whether the perturbed snippet is rejected based on either its names or its formatting.  
\stylish is not made aware what kind of perturbation (if any), identifier or whitespace,
was made to the snippet.  \autoref{fig:robustness}
reports \Projname's rejection performance as ROC curves. In each curve, each point corresponds to a different choice of threshold 
$T$, and the $x$-axis shows FPR (estimated as in \autoref{sec:fpr}), 
and the $y$-axis shows true positive rate, the proportion of the perturbed snippets that we correctly rejected.
 \Projname achieves high precision thus making it
suitable for use for a filtering pre-commit script.
For example, if the FPR is allowed to be at most $0.05$,  then we are able to correctly reject 40\% of the snippets.
The system is somewhat worse at rejecting snippets whose variable names have been perturbed; in part this is because predicting identifier names is more difficult than predicting formatting. 
New advances in language models for code \cite{maddison2014structured,nguyen2013statistical}
are likely to improve these results further.
Nonetheless, these results are promising: \stylish still rejects enough perturbed
snippets that if deployed at 5\% FPR, it 
would enhance convention adherence with minimal disruption to developers.

\subsection{Robustness of Suggestions}
\label{sec:eval:robustness}
We show that \Projname avoids two potential pitfalls in its identifier
suggestions: first, that it does not simply rename all tokens common ``junk''
names that appear in many contexts, and second, that it retains unusual names
that signify unusual functionality, adhering to the SUP.

\boldpara{Junk Names}
A junk name is a semantically uninformative name used in disparate contexts. 
It is difficult to formalize this concept:  for instance, in almost all cases,
 \lstinline{foo} and \lstinline{bar} are junk names, while \lstinline{i} 
 and \lstinline{j}, when used as loop counters, are semantically informative
  and therefore not junk. Despite this, most developers ``know it when they see it.’’
\begin{figure}[tb]
	\centering
	\includegraphics[width=.9\columnwidth]{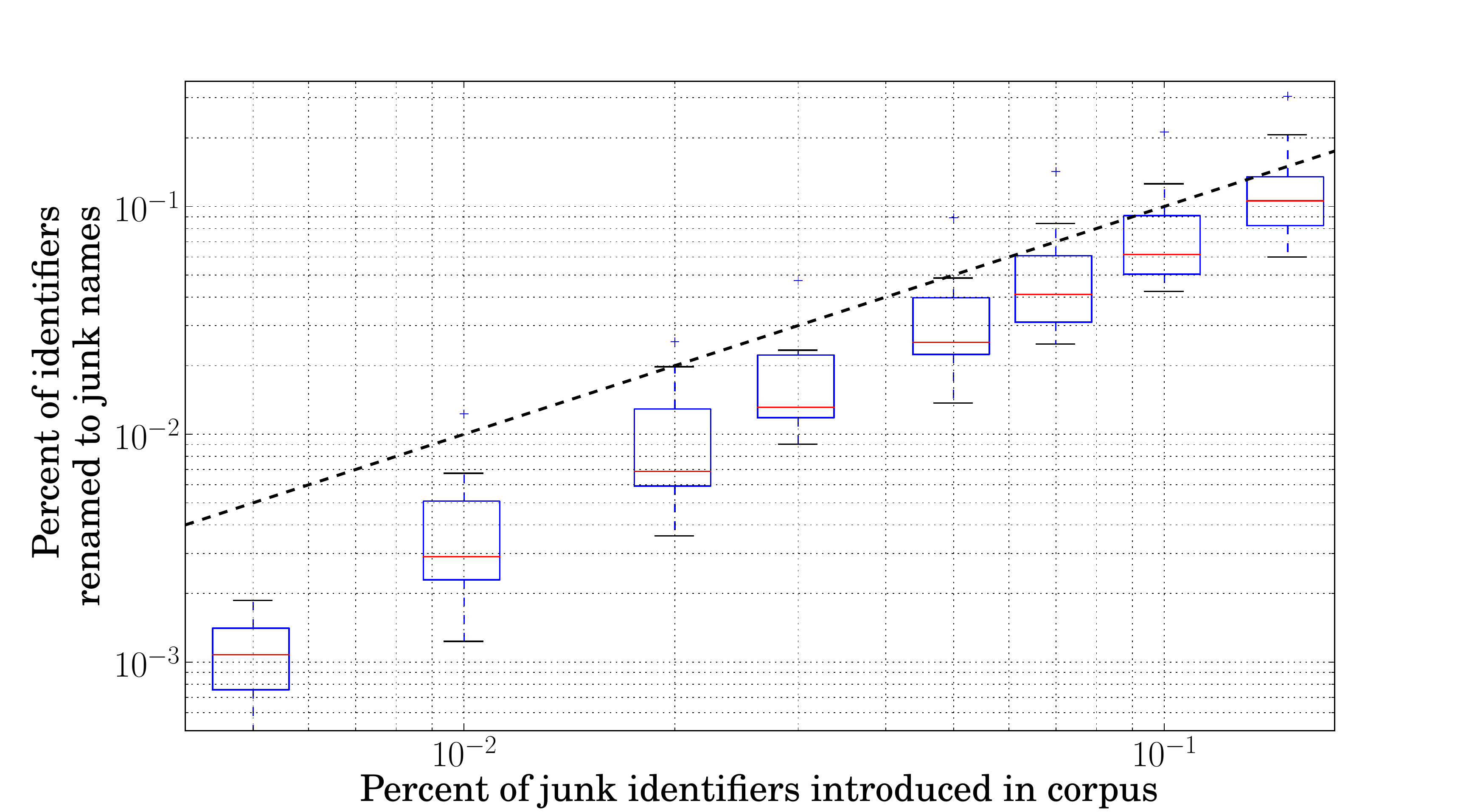}
	\caption{Is \Projname robust to low-quality corpora?
	The $x$-axis shows percentage of identifiers perturbed to junk
	names to simulate low quality corpus. The $y$-axis is percentage of 
	resulting low quality suggestions. Note log-log scale. 
	The dotted line shows $y=x$.  The boxplots are
	across the 10 evaluation projects.}
	\label{fig:junkEval}
\end{figure}
One might at first be concerned that \Projname would often suggest junk names,
because junk names appear in many different \ngrams in the training set.  We argue, however, that in fact 
the opposite
is the case:  \Projname actually \emph{resists}
suggesting junk names. This is because if a name appears in too many contexts,
it will be impossible to predict a unsurprising follow-up, and so code containing
junk names will have lower probability, and therefore worse score.

To evaluate this claim, we randomly rename variables to junk names in each
project to simulate a low quality project. Notice that we are simulating a low quality \emph{training} set,
which should be the worst case for \Projname.
We measure how our suggestions are affected by the proportion of junk names
in the training set. To generate junk
variables we use a discrete Zipf's law with slope $s=1.08$, the slope empirically
measured for all identifiers in our evaluation corpus. We verified the Zipfian  
assumption in previous work \cite{allamanis2013mining}.  \autoref{fig:junkEval} 
shows the effect on our suggestions 
 as the evaluation projects are gradually infected with
more junk names.  The framework successfully avoids
suggesting junk names, proposing them at a lower frequency than they exist in
the perturbed codebase.

\begin{figure}[t]
	\centering
	\includegraphics[width=.9\columnwidth]{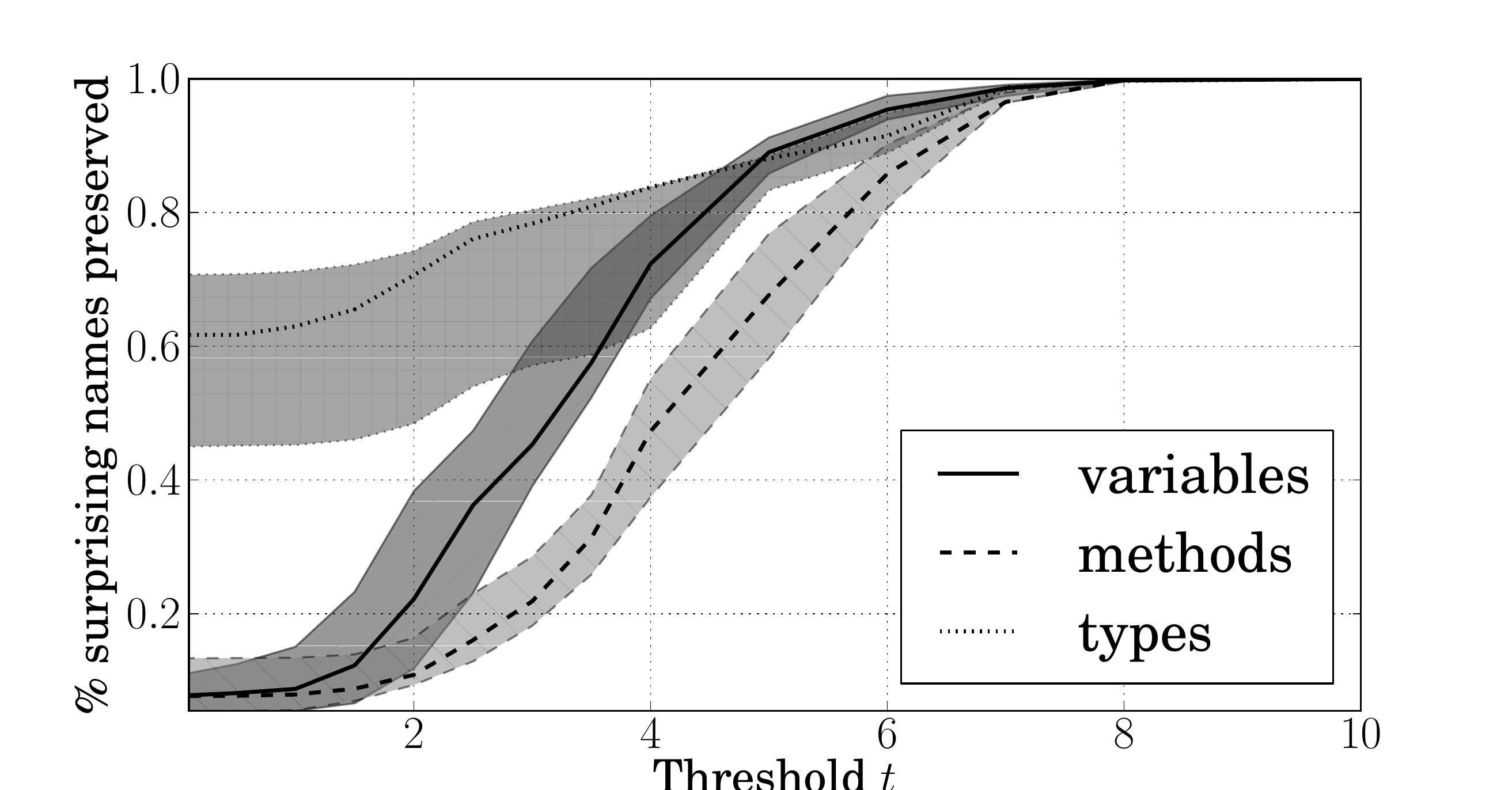}
  \caption{\Projname does not cause the ``heat death'' of a codebase: we
    evaluate the percent of single suggestions made on \UNK identifiers that
    preserve the surprising name. The $x$-axis shows the value of the threshold
    $t$, which controls the suggestion frequency of \suggest; lower $t$ means that \suggest
    has less freedom to decline to make low-quality suggestions.}
	\label{fig:heatdeathEval}
\end{figure}

\boldpara{Sympathetic Uniqueness} Surprise can be good in identifiers, where it signifies unusual
functionality.  Here we show that \Projname preserves this sort of surprise.
We find all identifiers in the test file
that are unknown to the LM, \ie are represented by an \UNK. We then plot the
percentage of those identifiers for which \suggest does \emph{not} propose an
alternative name, as a function of the threshold $t$.  As described in \autoref{sec:fpr}, 
$t$ is selected automatically, but it is useful to explore how adherence to the SUP
varies as a function of $t$. \autoref{fig:heatdeathEval} shows that
for reasonable threshold values, \Projname suggests non-\UNK identifiers for only a small proportion of
the \UNK identifiers (about 20\%). This confirms that \Projname does not cause
the ``heat death of a codebase'' by renaming semantically rich, surprising
names into frequent, low-content names.

\subsection{Manual Examination of Suggestions}
\label{sec:eval:needless}

As a qualitative evaluation of \Projname's suggestions,
three human
evaluators (three of the authors) independently evaluated 
the quality of its suggestions.  First, we
selected two projects from our corpus, uniformly at random, then for each we ran
\styleprofile on $30$ methods selected uniformly at random to
produce suggestions for all the identifiers present in that method.  
We assigned 20 profiles to each evaluator such
that each profile had two evaluators whose task was to \emph{independently}
determine whether any of the suggested renamings in a profile were reasonable.
The evaluators attempted to take an evidence based approach, that is,
not to simply choose names that they liked, but to choose names
that were consistent with existing practice in the project.
We provided access to the full
source code of each project to the evaluators.
 
One special issue arose when suggesting changes to method names.
Adapting linguistic terminology to our context, \emph{homonyms} are two
semantically equivalent functions with distinct names.  We deem \Projname's
suggested renaming of a method usage to be reasonable, if the suggestion list
contains a \emph{partial}, not necessarily perfect, homonym; \ie, if it draws the
developer's attention to another method that is used in similar contexts.  Each
evaluator had $15$ minutes to consider each profile and $30$ minutes to explore
each project before starting on a new project.   In fact, the evaluators
required much less time than allocated,
averaging about $5$ minutes per example.  The human evaluations can
be found on our project's webpage.
Surprisingly, given the quality of our evaluation codebases, $50$\% of 
the suggestions were determined to be useful by both evaluators.  Further,
no suggestion works well for everyone;  when we consider \Projname's performance
in terms of whether at least one evaluator found a suggestion useful, $63$\% of
the suggestions are useful,  with an inter-rater agreement (Cohen's kappa
\cite{carletta1996assessing}) of $\kappa=0.73$. 
The quality of suggestions is strikingly high given that these projects 
are mature, vibrant, high-profile projects.

This provides evidence that the naming suggestions provided by \Projname
are qualitatively reasonable.  Of course, an obvious threat to validity
is that the evaluators are not 
developers of the test projects, and the developers themselves 
may well have had different opinions about naming.
For this reason, we also provided the \Projname naming suggestions
to the project developers themselves, as described in the next section.


\subsection{Suggestions Accepted by Projects}
\label{sec:eval:patches}

We used \Projname's \styleprofile tool to identify high-confidence renamings and
submitted 18 of them as patches to the $5$ evaluation projects that actively use
GitHub. \autoref{tbl:projects} shows the pull request ids and their current
status. Four projects merged our pull requests (14 of 15 commits), while one other
ignored them without comment.  Developers in the projects that accepted
\Projname's patches found the \Projname useful: one said
\href{https://github.com/libgdx/libgdx/pull/1400}{``Wow, that's a pretty cool
tool!''}~\cite{libgdxPullRequest}. \junit did not accept two of the suggested
renamings as-is.  Instead, the patches sparked a discussion. Its developers
concluded that another name was more meaningful in one case and that the
suggested renaming of another violated the project's explicit naming convention:
``Renaming \texttt{e} to \texttt{t} is no improvement, because we should
consistently use \texttt{e}.''~\cite{junitPullRequest}. We then pointed them to
the code locations that supported \Projname's original suggestion. This triggered them to
change all the names that had caused the suggestion in the
first place --- which shows that \Projname pointed out an inconsistency, previously unnoticed, that improved the
naming in the project. Links to all the discussions can be found on
\href{http://groups.inf.ed.ac.uk/naturalize/}{our project webpage}.

\section{Related Work}
\label{sec:relwork}

Coding conventions, readability, and identifiers have been extensively studied in
the literature. Despite this, \Projname is --- to our knowledge
--- the first to infer coding conventions from a codebase, spanning
naming and formatting.

\textbf{Coding conventions} are standard
practice~\cite{Simula.SE.331,hatton2004safer}.  They facilitate consistent
measurement and reduce systematic error and generate more meaningful commits by
eliminating trivial convention enforcing commits
\cite{wikipedia:2013:codingconventions}.  Some programming languages like Java
and Python suggest specific coding
styles~\cite{oracle:codeconventions,rossum2008pep}, while consortia publish
guidelines for others, like C~\cite{motor2012misra,hatton2004safer}.

High quality \textbf{identifier names} lie at the heart of software engineering
\cite{anquetil1999recovering,brooks1975mythical,caprile:icsm:00,deiss:iwpc:05,Lawrie:sme:2007,soloway1984empirical,takang1996effects, ohba2005toward};
they drive code readability and comprehension~\cite{biggerstaff1993concept,buse2010learning, caprile:icsm:00,lawrie2006syntactic,liblit2006cognitive,takang:jpl:1996}.
According to Dei{\ss}enb{\"o}ck and Pizka \cite{broy2005holistic}, identifiers
represent the majority (70\%) of source code tokens. 
Eshkevari \etal~\cite{eshkevari2011exploratory} explored how
identifiers change in code, while Lawrie \etal
\cite{lawrie2006syntactic} studied the consistency of identifier namings.
Abebe \etal \cite{abebe2011effect} discuss the importance of
naming to concept location \cite{rajlich2004incremental}. 
Caprile and Tonella \cite{caprile:icsm:00} propose a 
framework for restructuring and renaming identifiers based on
custom rules and dictionaries. 
Gupta \etal present part-of-speech tagging on split multi-word identifiers to improve
software engineering tools~\cite{gupta2013part}.
H{\o}st and {\O}stvold stem and tag method names, then learn a 
mapping from them to a fixed set of predicates over bytecode.  Naming bugs are mismatches
between the map and stemmed and tagged method names and their predicates in a
test set~\cite{host09debuggingmethod}.
In contrast, our work considers coding conventions
more generally, and takes a flexible, data-driven approach.
 Several styles exist for
engineering consistent identifiers
\cite{binkley:icpc:09,caprile:icsm:00,deiss:iwpc:05,simonyi:msdn:99}. Because
longer names are more
informative~\cite{liblit2006cognitive}, these styles share an agglutination
mechanism for creating multi-word names~\cite{anquetil1998assessing,ratiu2007reality}.  

Many rule-based \textbf{code formatters} exist but, to our knowledge, 
are limited to constraining identifier names to obey constraints like
CamelCase or underscore and cannot handle convention like 
the use of \lstinline+i+ as a loop control variable. 
\texttt{Pylint} \cite{pylint} checks if names match a simple set of 
regular expressions (\eg, variable names must be lowercase);
\texttt{astyle}, \texttt{aspell}
and GNU \texttt{indent} \cite{astyle, gnuindent} only format whitespace tokens.
\texttt{gofmt} formats the code aiming to ``eliminating an entire class of argument''
among developers \cite[slide 66]{gofmt} but provide no guidance
for naming. Wang \etal have developed a heuristic-based method to automatically
insert blank lines into methods (vertical spacing) to improve readability~\cite{wang2011automatic}.
\Projname is unique in that it does not require upfront agreement on
hard rules but learns soft rules that are implicit in a codebase.
The soft rules about which \Projname is highly confident can be extracted for
by a formatter.

\textbf{API recommenders, code suggestion and completion systems} aim to help during
 editing, when a user may not know the name of an API she
needs~\cite{robillard2010recommendation} or that the API call she is making
needs to preceded by
another~\cite{gabel2011inferring,uddin2011analyzing,wang2013mining,zhong2009mapo}.
Code suggestion and completion tools \cite{bruch09learning,hindle2012naturalness, nguyen12graph, nguyen2013statistical, robbes08program} suggest the next token 
during editing, often from a few initial characters.
Essentially these methods address a search problem, helping the
developer find an existing entity in code.
Our focus is instead on release management and improving code's adherence to convention.
For example, code completion engines will not suggest renaming parameters like \lstinline+str+ in \autoref{fig:msex}.

Language models are extensively used
in \textbf{natural language processing}, especially in speech recognition and machine
translation \cite{chen1996empirical,jurafsky:martin}. Despite this
extensive work, LMs have been under-explored for non-ambiguous (\eg programming) languages,
with only a few recent exceptions~\cite{allamanis2013mining,hindle2012naturalness,maddison2014structured,nguyen2013statistical}.
The probabilistic nature of language models allows us to
tackle the suggestion problem in a principled way.  
There is very little work on using NLP tools to suggest revisions to improve
existing text. The main exception is spelling correction \cite{kukich92spelling},
to which LMs have been applied \cite{mays91context}.  However, spelling correction
methods often rely on strong assumptions about what errors are
most common, a powerful source of information which has no analog
in our domain.

\section{Conclusion}
\label{sec:conc}

We have presented \Projname, the first tool that learns local style
from a codebase and provides suggestions to improve stylistic consistency.
We have taken the view that conventions are a matter of \emph{mores} rather
than \emph{laws}: We suggest changes only when there is sufficient
evidence of emerging consensus in the codebase.
We showed that \Projname is effective at making natural suggestions,
achieving 94\% accuracy in its top suggestions for identifier names,
and even suggesting useful revisions to mature, high quality open source projects.

\cut{
Future work could include integrating in recent language models for code \cite{maddison2014structured,nguyen2013statistical}
and integrating \Projname into a review tool like Gerrit~\cite{gerrit}.
We suspect that exploring other methods of incorporating syntactic context
could further improve performance.  A broad empirical study
of local conventions could be of interest, to explore
whether some conventions are more effective than others, or if what
is most important is simply to pick one.
}

\cut{
The question of whether *any* code conventions can be shown to be superior to
others would be an interesting empirical study of its own, but probably tedious
since it would perforce be a user study:  useful in our context for speeding
convergence whenever such a result can be shown.  To side-step a user study,
could we show superiority in terms of cross-entropy?  The governing intuition
here is that things that should draw a developer's attention should have higher 
cross-entropy, while boilerplate should fade into the background and have lower
cross-entropy.
}


\section*{Acknowledgements}
This work was supported by Microsoft Research through
its PhD Scholarship Programme and the Engineering and Physical Sciences
Research Council [grant number EP/K024043/1].
We acknowledge Mehrdad Afshari who first asked us about 
junk variables and the heat death of the codebase.

\bibliographystyle{abbrv}
\balance
\bibliography{lit/learningCodingConventions}

\begin{thebibliography}{10}

\bibitem{abebe2011effect}
S.~L. Abebe, S.~Haiduc, P.~Tonella, and A.~Marcus.
\newblock The effect of lexicon bad smells on concept location in source code.
\newblock In {\em Source Code Analysis and Manipulation (SCAM), 2011 11th IEEE
  International Working Conference on}, pages 125--134. IEEE, 2011.

\bibitem{swebok:2004}
A.~Abran, P.~Bourque, R.~Dupuis, J.~W. Moore, and L.~L. Tripp.
\newblock {\em {Guide to the Software Engineering Body of Knowledge - SWEBOK}}.
\newblock IEEE Press, Piscataway, NJ, USA, 2004 version edition, 2004.

\bibitem{adams:ibm:1984}
E.~N. Adams.
\newblock Optimizing preventive service of software products.
\newblock {\em IBM Journal of Research and Development}, 28(1):2--14, Jan.
  1984.

\bibitem{allamanis2013mining}
M.~Allamanis and C.~Sutton.
\newblock Mining source code repositories at massive scale using language
  modeling.
\newblock In {\em Proceedings of the Tenth International Workshop on Mining
  Software Repositories}, pages 207--216. IEEE Press, 2013.

\bibitem{anquetil1998assessing}
N.~Anquetil and T.~Lethbridge.
\newblock Assessing the relevance of identifier names in a legacy software
  system.
\newblock In {\em Proceedings of the 1998 conference of the Centre for Advanced
  Studies on Collaborative Research}, page~4, 1998.

\bibitem{anquetil1999recovering}
N.~Anquetil and T.~C. Lethbridge.
\newblock Recovering software architecture from the names of source files.
\newblock {\em Journal of Software Maintenance}, 11(3):201--221, 1999.

\bibitem{arthur2014apple}
C.~Arthur.
\newblock {Apple's SSL iPhone vulnerability: how did it happen, and what next?}
\newblock \href{http://bit.ly/1bJ7aSa}{\texttt{bit.ly/1bJ7aSa}}, 2014.
\newblock Visited Mar 2014.

\bibitem{motor2012misra}
M.~I. S.~R. Association et~al.
\newblock {MISRA-C 2012: Guidelines for the Use of the C Language in Critical
  Systems}.
\newblock {\em ISBN 9781906400118}, 2012.

\bibitem{astyle}
astyle Contributors.
\newblock Artistic style 2.03.
\newblock \url{http://astyle.sourceforge.net/}, 2013.
\newblock Visited September 9, 2013.

\bibitem{bacchelli2013eoc}
A.~Bacchelli and C.~Bird.
\newblock {Expectations, Outcomes, and Challenges of Modern Code Review}.
\newblock In {\em ICSE}, 2013.

\bibitem{bergstra2012random}
J.~Bergstra and Y.~Bengio.
\newblock Random search for hyper-parameter optimization.
\newblock {\em The Journal of Machine Learning Research}, 13:281--305, 2012.

\bibitem{biggerstaff1993concept}
T.~J. Biggerstaff, B.~G. Mitbander, and D.~Webster.
\newblock The concept assignment problem in program understanding.
\newblock In {\em Proceedings of the 15th international conference on Software
  Engineering}, pages 482--498. IEEE Computer Society Press, 1993.

\bibitem{binkley:emse:13}
D.~Binkley, M.~Davis, D.~Lawrie, J.~Maletic, C.~Morrell, and B.~Sharif.
\newblock The impact of identifier style on effort and comprehension.
\newblock {\em Empirical Software Engineering}, 18(2):219--276, 2013.

\bibitem{binkley:icpc:09}
D.~Binkley, M.~Davis, D.~Lawrie, and C.~Morrell.
\newblock To {CamelCase} or {Under\_score}.
\newblock In {\em IEEE 17th International Conference on Program Comprehension
  ({ICPC}'09)}, pages 158--167, 2009.

\bibitem{Simula.SE.331}
C.~Boogerd and L.~Moonen.
\newblock Assessing the value of coding standards: An empirical study.
\newblock In H.~Mei and K.~Wong, editors, {\em Proceedings of the 24th IEEE
  International Conference on Software Maintenance (ICSM 2008)}, pages 277 --
  286. IEEE, October 2008.

\bibitem{brooks1975mythical}
F.~P. Brooks.
\newblock {\em The mythical man-month}.
\newblock Addison-Wesley Reading, 1975.

\bibitem{broy2005holistic}
M.~Broy, F.~Dei{\ss}enb{\"o}ck, and M.~Pizka.
\newblock A holistic approach to software quality at work.
\newblock In {\em Proc. 3rd World Congress for Software Quality (3WCSQ)}, 2005.

\bibitem{bruch09learning}
M.~Bruch, M.~Monperrus, and M.~Mezini.
\newblock Learning from examples to improve code completion systems.
\newblock In {\em ESEC/SIGSOFT FSE}, pages 213--222. ACM, 2009.

\bibitem{buse2010learning}
R.~P. Buse and W.~R. Weimer.
\newblock Learning a metric for code readability.
\newblock {\em Software Engineering, IEEE Transactions on}, 36(4):546--558,
  2010.

\bibitem{caprile:icsm:00}
B.~Caprile and P.~Tonella.
\newblock Restructuring program identifier names.
\newblock In {\em International Conference on Software Maintenance
  ({ICSM}'00)}, pages 97--107, 2000.

\bibitem{carletta1996assessing}
J.~Carletta.
\newblock Assessing agreement on classification tasks: the kappa statistic.
\newblock {\em Computational linguistics}, 22(2):249--254, 1996.

\bibitem{chen1996empirical}
S.~Chen and J.~Goodman.
\newblock An empirical study of smoothing techniques for language modeling.
\newblock In {\em Proceedings of the 34th annual meeting on Association for
  Computational Linguistics}, pages 310--318. Association for Computational
  Linguistics, 1996.

\bibitem{cowan2001magical}
N.~Cowan.
\newblock The magical number 4 in short-term memory: A reconsideration of
  mental storage capacity.
\newblock {\em Behavioral and Brain Sciences}, 24(1):87--114, 2001.

\bibitem{deiss:iwpc:05}
F.~Dei{\ss}enb\"ock and M.~Pizka.
\newblock Concise and consistent naming [software system identifier naming].
\newblock In {\em Proceedings of the 13th International Workshop on Program
  Comprehension (IWPC'05)}, pages 97--106, 2005.

\bibitem{dowdy2011statistics}
S.~Dowdy, S.~Wearden, and D.~Chilko.
\newblock {\em Statistics for research}, volume 512.
\newblock John Wiley \& Sons, 2011.

\bibitem{eclipseJdt}
Eclipse-Contributors.
\newblock Eclipse {JDT}.
\newblock \url{http://www.eclipse.org/jdt/}, 2013.
\newblock Visited September 9, 2013.

\bibitem{eshkevari2011exploratory}
L.~M. Eshkevari, V.~Arnaoudova, M.~Di~Penta, R.~Oliveto, Y.-G.
  Gu{\'e}h{\'e}neuc, and G.~Antoniol.
\newblock An exploratory study of identifier renamings.
\newblock In {\em Proceedings of the 8th Working Conference on Mining Software
  Repositories}, pages 33--42. ACM, 2011.

\bibitem{gabel:fse:2010}
M.~Gabel and Z.~Su.
\newblock A study of the uniqueness of source code.
\newblock In {\em Proceedings of the eighteenth ACM SIGSOFT international
  symposium on Foundations of software engineering}, FSE '10, pages 147--156,
  New York, NY, USA, 2010. ACM.

\bibitem{gabel2011inferring}
M.~G. Gabel.
\newblock {\em Inferring Programmer Intent and Related Errors from Software}.
\newblock PhD thesis, University of California, 2011.

\bibitem{junitPullRequest}
GitHub.
\newblock {JUnit Pull Request \#834}.
\newblock \href{http://bit.ly/O8bmjM}{\texttt{bit.ly/O8bmjM}}, 2014.
\newblock Visited Mar 2014.

\bibitem{libgdxPullRequest}
GitHub.
\newblock {libgdx Pull Request \#1400}.
\newblock \href{http://bit.ly/O8aBqV}{\texttt{bit.ly/O8aBqV}}, 2014.
\newblock Visited Mar 2014.

\bibitem{gnuindent}
gnu-indent Contributors.
\newblock {GNU Indent -- beautify C code}.
\newblock \url{http://www.gnu.org/software/indent/}, 2013.
\newblock Visited September 9, 2013.

\bibitem{gupta2013part}
S.~Gupta, S.~Malik, L.~Pollock, and K.~Vijay-Shanker.
\newblock Part-of-speech tagging of program identifiers for improved text-based
  software engineering tools.
\newblock In {\em International Conference on Program Comprehension}, pages
  3--12. IEEE, 2013.

\bibitem{hatton2004safer}
L.~Hatton.
\newblock Safer language subsets: an overview and a case history, {MISRA C}.
\newblock {\em Information and Software Technology}, 46(7):465--472, 2004.

\bibitem{hindle2012naturalness}
A.~Hindle, E.~T. Barr, Z.~Su, M.~Gabel, and P.~Devanbu.
\newblock On the naturalness of software.
\newblock In {\em Software Engineering (ICSE), 2012 34th International
  Conference on}, pages 837--847. IEEE, 2012.

\bibitem{hindle:scp:2009}
A.~Hindle, M.~W. Godfrey, and R.~C. Holt.
\newblock Reading beside the lines: Using indentation to rank revisions by
  complexity.
\newblock {\em Science of Computer Programming}, 74(7):414--429, May 2009.

\bibitem{host09debuggingmethod}
E.~W. H{\o}st and B.~M. {\O}stvold.
\newblock Debugging method names.
\newblock In {\em In European Conference on Object-Oriented Programming
  (ECOOP}, pages 294--317. Springer, 2009.

\bibitem{jurafsky:martin}
D.~Jurafsky and J.~H. Martin.
\newblock {\em Speech and Language Processing: An Introduction to Natural
  Language Processing, Computational Linguistics and Speech Recognition}.
\newblock Prentice Hall, 2nd edition, 2009.

\bibitem{kukich92spelling}
K.~Kukich.
\newblock Techniques for automatically correcting words in text.
\newblock {\em ACM Computing Surveys}, 24(4):377--439, Dec. 1992.

\bibitem{langley2014bug}
A.~Langley.
\newblock {Apple's SSL/TLS bug}.
\newblock \href{http://bit.ly/MMvx6b}{\texttt{bit.ly/MMvx6b}}, 2014.
\newblock Visited Mar 2014.

\bibitem{lawrie2006syntactic}
D.~Lawrie, H.~Feild, and D.~Binkley.
\newblock Syntactic identifier conciseness and consistency.
\newblock In {\em Source Code Analysis and Manipulation, 2006. SCAM'06. Sixth
  IEEE International Workshop on}, pages 139--148. IEEE, 2006.

\bibitem{Lawrie:sme:2007}
D.~Lawrie, H.~Feild, and D.~Binkley.
\newblock An empirical study of rules for well-formed identifiers: Research
  articles.
\newblock {\em Journal of Software Maintenance Evolution: Research and
  Practice}, 19(4):205--229, July 2007.

\bibitem{lawrie:icpc:2006}
D.~Lawrie, C.~Morrell, H.~Feild, and D.~Binkley.
\newblock {What's in a Name? A Study of Identifiers}.
\newblock In {\em Proceedings of the 14th IEEE International Conference on
  Program Comprehension ({ICPC}'06)}, ICPC '06, pages 3--12, Washington, DC,
  USA, 2006. IEEE Computer Society.

\bibitem{liblit2006cognitive}
B.~Liblit, A.~Begel, and E.~Sweetser.
\newblock Cognitive perspectives on the role of naming in computer programs.
\newblock In {\em Annual Psychology of Programming Workshop}, 2006.

\bibitem{rouge}
C.-Y. Lin.
\newblock Rouge: A package for automatic evaluation of summaries.
\newblock In {\em Text Summarization Branches Out: Proceedings of the ACL-04
  Workshop}, pages 74--81, 2004.

\bibitem{maddison2014structured}
C.~J. Maddison and D.~Tarlow.
\newblock Structured generative models of natural source code.
\newblock {\em arXiv preprint arXiv:1401.0514}, 2014.

\bibitem{mays91context}
E.~Mays, F.~J. Damerau, and R.~L. Mercer.
\newblock Context based spelling correction.
\newblock {\em Information Processing and Management}, 27(5):517--522, 1991.

\bibitem{miller1956magical}
G.~A. Miller.
\newblock The magical number seven, plus or minus two: some limits on our
  capacity for processing information.
\newblock {\em Psychological review}, 63(2):81, 1956.

\bibitem{movshovitz2013natural}
D.~Movshovitz-Attias and W.~W.~Cohen.
\newblock Natural language models for predicting programming comments.
\newblock In {\em Proc of the ACL}, 2013.

\bibitem{murphy2009refactor}
E.~Murphy-Hill, C.~Parnin, and A.~P. Black.
\newblock How we refactor, and how we know it.
\newblock {\em Software Engineering, IEEE Transactions on}, 38(1):5--18, 2012.

\bibitem{nagappan:esem:07}
N.~Nagappan and T.~Ball.
\newblock Using software dependencies and churn metrics to predict field
  failures: An empirical case study.
\newblock In {\em ESEM}, pages 364--373, 2007.

\bibitem{nguyen12graph}
A.~T. Nguyen, T.~T. Nguyen, H.~A. Nguyen, A.~Tamrawi, H.~V. Nguyen,
  J.~Al-Kofahi, and T.~N. Nguyen.
\newblock Graph-based pattern-oriented, context-sensitive source code
  completion.
\newblock In {\em Proceedings of the 34th ACM/IEEE International Conference on
  Software Engineering (ACM/IEEE ICSE 2012)}. IEEE, 2012.

\bibitem{nguyen2013statistical}
T.~T. Nguyen, A.~T. Nguyen, H.~A. Nguyen, and T.~N. Nguyen.
\newblock A statistical semantic language model for source code.
\newblock In {\em Proceedings of the 2013 9th Joint Meeting on Foundations of
  Software Engineering}, pages 532--542. ACM, 2013.

\bibitem{ohba2005toward}
M.~Ohba and K.~Gondow.
\newblock Toward mining concept keywords from identifiers in large software
  projects.
\newblock In {\em ACM SIGSOFT Software Engineering Notes}, volume~30, pages
  1--5. ACM, 2005.

\bibitem{oracle:codeconventions}
Oracle.
\newblock {Code Conventions for the Java Programming Language}.
\newblock \url{http://www.oracle.com/technetwork/java/codeconv-138413.html},
  1999.
\newblock Visited September 2, 2013.

\bibitem{bleu}
K.~Papineni, S.~Roukos, T.~Ward, and W.-J. Zhu.
\newblock {BLEU:} a method for automatic evaluation of machine translation.
\newblock In {\em Association for Computational Linguistics (ACL)}, pages
  311--318, 2002.

\bibitem{gofmt}
R.~Pike.
\newblock Go at {G}oogle.
\newblock \url{http://talks.golang.org/2012/splash.slide}, 2012.
\newblock Visited September 9, 2013.

\bibitem{pylint}
Pylint-Contributors.
\newblock Pylint -- code analysis for {P}ython.
\newblock \url{http://www.pylint.org/}, 2013.
\newblock Visited September 9, 2013.

\bibitem{rajlich2004incremental}
V.~Rajlich and P.~Gosavi.
\newblock Incremental change in object-oriented programming.
\newblock {\em Software, IEEE}, 21(4):62--69, 2004.

\bibitem{ratiu2007reality}
D.~Ratiu and F.~Dei{\ss}enb{\"o}ck.
\newblock From reality to programs and (not quite) back again.
\newblock In {\em Program Comprehension, 2007. ICPC'07. 15th IEEE International
  Conference on}, pages 91--102. IEEE, 2007.

\bibitem{rigby2013convergent}
P.~C. Rigby and C.~Bird.
\newblock Convergent software peer review practices.
\newblock In {\em Proceedings of the the joint meeting of the European Software
  Engineering Conference and the ACM SIGSOFT Symposium on The Foundations of
  Software Engineering (ESEC/FSE)}. ACM, 2013.

\bibitem{robbes08program}
R.~Robbes and M.~Lanza.
\newblock How program history can improve code completion.
\newblock In {\em Automated Software Engineering (ASE)}, pages 317--326. IEEE,
  2008.

\bibitem{robillard2010recommendation}
M.~Robillard, R.~Walker, and T.~Zimmermann.
\newblock Recommendation systems for software engineering.
\newblock {\em Software, IEEE}, 27(4):80--86, 2010.

\bibitem{rossum2008pep}
G.~v. Rossum, B.~Warsaw, and N.~Coghlan.
\newblock {PEP 8--Style Guide for Python Code}.
\newblock \url{http://www.python.org/dev/peps/pep-0008/}, 2013.
\newblock Visited September 8, 2013.

\bibitem{simonyi:msdn:99}
C.~Simonyi.
\newblock Hungarian notation.
\newblock \url{http://msdn.microsoft.com/en-us/library/aa260976(VS.60).aspx},
  1999.
\newblock Visited September 2, 2013.

\bibitem{soloway1984empirical}
E.~Soloway and K.~Ehrlich.
\newblock Empirical studies of programming knowledge.
\newblock {\em Software Engineering, IEEE Transactions on}, (5):595--609, 1984.

\bibitem{strunk:white}
W.~Strunk~Jr and E.~White.
\newblock {\em The elements of style}.
\newblock Macmillan, New York, 3rd edition, 1979.

\bibitem{takang:jpl:1996}
A.~Takang, P.~Grubb, and R.~Macredie.
\newblock The effects of comments and identifier names on program
  comprehensibility: an experiential study.
\newblock 4(3):143--167, 1996.

\bibitem{takang1996effects}
A.~A. Takang, P.~A. Grubb, and R.~D. Macredie.
\newblock The effects of comments and identifier names on program
  comprehensibility: an experimental investigation.
\newblock {\em J. Prog. Lang.}, 4(3):143--167, 1996.

\bibitem{uddin2011analyzing}
G.~Uddin, B.~Dagenais, and M.~P. Robillard.
\newblock Analyzing temporal {API} usage patterns.
\newblock In {\em Proceedings of the 2011 26th IEEE/ACM International
  Conference on Automated Software Engineering}, pages 456--459. IEEE Computer
  Society, 2011.

\bibitem{wang2013mining}
J.~Wang, Y.~Dang, H.~Zhang, K.~Chen, T.~Xie, and D.~Zhang.
\newblock Mining succinct and high-coverage {API} usage patterns from source
  code.
\newblock In {\em Proceedings of the Tenth International Workshop on Mining
  Software Repositories}, pages 319--328. IEEE Press, 2013.

\bibitem{wang2011automatic}
X.~Wang, L.~Pollock, and K.~Vijay-Shanker.
\newblock Automatic segmentation of method code into meaningful blocks to
  improve readability.
\newblock In {\em Working Conference on Reverse Engineering}, pages 35--44.
  IEEE, 2011.

\bibitem{wikipedia:2013:codingconventions}
{W}ikipedia.
\newblock Coding {C}onventions.
\newblock \url{http://en.wikipedia.org/wiki/Coding_conventions}.

\bibitem{young1996economics}
H.~P. Young.
\newblock The economics of convention.
\newblock {\em The Journal of Economic Perspectives}, 10(2):105--122, 1996.

\bibitem{zhong2009mapo}
H.~Zhong, T.~Xie, L.~Zhang, J.~Pei, and H.~Mei.
\newblock {MAPO: Mining and recommending API usage patterns}.
\newblock In {\em ECOOP 2009--Object-Oriented Programming}, pages 318--343.
  Springer, 2009.

\end{thebibliography}

\end{document}